\documentclass[aps,10pt,pra,superscriptaddress,showpacs,floatfix,twocolumn,longbibliography,nofootinbib]{revtex4-1}

\AtBeginDocument{%
  \providecommand\BibTeX{{%
    \normalfont B\kern-0.5em{\scshape i\kern-0.25em b}\kern-0.8em\TeX}}}

\usepackage{listings}
\usepackage{graphicx}
\usepackage{dcolumn}
\usepackage{bm}
\usepackage{dsfont}
\usepackage{physics}
\usepackage[toc, page]{appendix}
\usepackage{epstopdf}
\usepackage{comment}
\usepackage{hyperref}
\usepackage{subfigure}

\newcommand{\beq}{\begin{equation}}
\newcommand{\eeq}{\end{equation}}
\newcommand{\beqnn}{\begin{equation*}}
\newcommand{\eeqnn}{\end{equation*}}
\newcommand{\bea}{\begin{eqnarray}}
\newcommand{\eea}{\end{eqnarray}}
\newcommand{\beann}{\begin{eqnarray*}}
\newcommand{\eeann}{\end{eqnarray*}}
\newcommand{\bes} {\begin{subequations}}
\newcommand{\ees} {\end{subequations}}

\newcommand{\ignore}[1]{}

\begin{document}

\title{Testing a quantum annealer as a quantum thermal sampler}

\author{Zoe Gonzalez Izquierdo}
\affiliation{QuAIL, NASA Ames Research Center, USA and USRA Research Institute for Advanced Computer Science, Mountain View, California 94043}

\author{Tameem Albash}
\affiliation{Department of Electrical and Computer Engineering,  University of New Mexico, Albuquerque, New Mexico 87131, USA}
\affiliation{Department of Physics and Astronomy and Center for Quantum Information and Control, CQuIC, University of New Mexico, Albuquerque, New Mexico 87131, USA}

\author{Itay Hen}
\email{itayhen@isi.edu}
\affiliation{Department of Physics and Astronomy, and Center for Quantum Information Science \& Technology, University of Southern California, Los Angeles, California 90089, USA}
\affiliation{Information Sciences Institute, University of Southern California, Marina del Rey, California 90292, USA}

\begin{abstract}
  Motivated by recent experiments in which specific thermal properties of complex many-body systems were successfully reproduced on a commercially available quantum annealer, we examine the extent to which quantum annealing hardware can reliably sample from the thermal state in a specific basis associated with a target quantum Hamiltonian.  We address this question by studying the diagonal thermal properties of the canonical one-dimensional transverse-field Ising model on a D-Wave 2000Q quantum annealing processor. We find that the quantum processor fails to produce the correct expectation values predicted by Quantum Monte Carlo. Comparing to master equation simulations, we find that this discrepancy is best explained by how the measurements at finite transverse fields are enacted on the device.  Specifically, measurements at finite transverse field require the system to be quenched from the target Hamiltonian to a Hamiltonian with negligible transverse field, and this quench is too slow.  The limitations imposed by such hardware make it an unlikely candidate for thermal sampling, and it remains an open question what thermal expectation values can be robustly estimated in general for arbitrary quantum many-body systems.
\end{abstract}




\maketitle

\section{Introduction}\label{sec:intro}
Describing the static and dynamical properties of a many-body quantum system remains a considerable challenge in physics.  Strong correlations among the particles of a many-body quantum system can lead to highly complex entangled states, which exist in a vector space whose dimension grows exponentially with the size of the system. Due to this exponential scaling, affecting both the time and memory that a classical computer needs in order to perform the relevant computations, even simulations of systems with only a few correlated particles quickly become intractable.  Tackling this problem with a quantum computer or simulator continues to inspire~\cite{Feynman1981} and drive the field of quantum simulation~\cite{Lloyd:96,Cirac:2012pi,Childs2018,Cao2018}.  Nevertheless, current analog quantum simulators are still limited by decoherence and implementation errors, and it remains an open question to identify  ``accessible properties of quantum systems which are robust with respect to error, yet are also hard to simulate classically'' \cite{Preskill2018quantumcomputingin}. 

The adiabatic paradigm of quantum computing~\cite{Farhi:00,aharonov_adiabatic_2007,AharonovTa-Shma} naturally lends itself to tackling the problem of studying ground state properties of quantum systems.  By preparing the system in the ground state of a `trivial' Hamiltonian and performing a sufficiently slow interpolation towards the target many-body quantum Hamiltonian, the adiabatic theorem of quantum mechanics~\cite{born_beweis_1928,kato_adiabatic_1950,Jansen:07} provides a guarantee that the state at the end of the evolution will be close to the target ground state.  

But what if we are interested in finite temperature properties of the same system? For decades, the work-horse for addressing this question has been Quantum Monte Carlo (QMC) methods~\cite{qmc1986,Landau:2005:GMC:1051461,newman}. Despite extensive work on developing more sophisticated methods~\cite{sandvik:92,prokofiev:98,sandvik:99,ODE}, no universal method whose space and time complexity scale polynomially with problem size is yet known. It remains an open research topic how to efficiently encode a generic thermal state into the ground state of a quantum Hamiltonian~\cite{PhysRevLett.99.030603,Nishimori:2015dp}.  

Alternatively, one can ask whether the open-system dynamics~\cite{Breuer:2002} of a system naturally has the thermal state as its steady state, and if so whether the dissipative dynamics can be used in lieu of a true quantum algorithm to prepare such a state. Having the steady state of the dissipative dynamics be the standard Gibbs state associated with the target Hamiltonian is a non-trivial assumption; it is known to be the case of Markovian weak-coupling limit master equations~\cite{Davies:74,ABLZ:12-SI} satisfying the Kubo-Martin-Schwinger (KMS) condition~\cite{KMS}. An open-system adiabatic theorem~\cite{PhysRevA.82.040304,salem_quasi-static_2007,joye_general_2007,springerlink:10.1007/BF01011696,PhysRevA.93.032118} provides a guarantee that in the long-time limit the state of the system will be close to the desired thermal state.

Recently, experiments~\cite{King2018,Harris2018,King2019,Kai2020} carried out on D-Wave quantum annealing processors were reported to have successfully reproduced certain thermal properties of complex quantum systems. In these studies, the processor is used as an analog quantum simulator: the superconducting circuit hardware~\cite{harris_flux_qubit_2010,Johnson:2010ys,Berkley:2010zr,Bunyk:2014hb} implements a transverse-field Ising model (TFIM) Hamiltonian in its low-lying spectrum.  The system is allowed to thermally relax at a particular realization of the TFIM Hamiltonian, and diagonal thermodynamic observables are calculated using measurements performed after quenching the system to a purely Ising Hamiltonian.  In Refs.~\cite{King2018,King2019}, the emergence of the finite order parameter associated with the Kosterlitz-Thouless phase transition of the TFIM on the triangular lattice~\cite{PhysRevB.68.104409} was observed; Ref.~\cite{Harris2018} observed the finite-size precursors of the phase transitions associated with the TFIM on three-dimensional cubic lattices, and Ref.~\cite{Kai2020} observed the phase transitions from the N{\'e}el antiferromagnetic phase to the 1/3 plateau.
These studies found good agreement between theory and experimental observations when measured quantities were restricted to the order parameter of the studied system~\cite{King2018,King2019} or to critical parameters indicating the location of phase transitions~\cite{Harris2018}.  (The results of Ref.~\cite{Kai2020} exhibit smoothed out results for the order parameter instead of the sharp transitions expected from theory.)

Motivated by these results, we examine in this study the extent to which one can use such hardware to provide thermal samples in a given basis for quantum Hamiltonians implemented by the hardware, especially at ever increasing system sizes.  Such samples, similar to those generated by QMC algorithms, would allow us to calculate the thermal expectation values of observables that are diagonal in this basis. As the simulation of quantum systems on quantum information processors gains traction and moves on to problems which are classically difficult (or virtually impossible) to solve, we face the question of how to verify or trust the results produced by these devices.  The one-dimensional (1D) TFIM, a canonical system in quantum thermodynamics, provides a simple case study for the question of thermal state sampling, which hopefully helps build confidence in the device as more difficult problems are tackled. 

To that aim, we study the performance of the D-Wave 2000Q annealing processor (DW)\footnote{Housed at NASA Ames Research Center.} on the 1D nonfrustrated TFIM. We verify the accuracy of the expectation values calculated from the experimental samples by comparing them to data obtained from QMC simulations. If such devices are to be used more widely as thermal state samplers, we believe this is a crucial first step in the process of validating the reliability of the device for this task.

We find in general that the thermal expectation values calculated using the samples from the annealer are not in good agreement with the expected values obtained by QMC, especially in the quantum paramagnetic region where the transverse field dominates~\cite{Sachdev_QPT}.  Our results are consistent with other findings that the devices samples are not thermal \cite{King2018,King2019}, and for the 1D TFIM the diagonal observables we study are too sensitive to the discrepancy between the measured and thermal distributions.
Master equation simulations~\cite{Albash2012} on small size problems confirm that a reason for this discrepancy is the inability of current devices to perform measurements at the target Hamiltonian; instead measurements are performed on an Ising Hamiltonian with negligible transverse field, which requires a quench from the target Hamiltonian to the Ising Hamiltonian.
Our simulations indicate that increasing the maximum annealing rate by a factor of $10^5$ would reproduce the expected thermal expectation values, but this would in turn require annealing times on the order of a few picoseconds. Extensivity of the Hamiltonian suggests that this annealing rate would need to decrease with the inverse of the system size.

\section{Experimental Methods}
\label{sec:methods}

\subsection{Transverse-Field Ising Model}
\label{subsec:TFIM}

We consider the one-dimensional transverse-field Ising model (1D TFIM)~\cite{Sachdev_QPT} with the $n$-qubit  Hamiltonian
\begin{equation} \label{eqt:H}
H_{\mathrm{TFIM}} = -A H_{\mathrm{TF}} + B H_{\mathrm{IM}} =  -  A \sum_{i=1}^n \sigma^x_i - B \sum_{i=1}^n \sigma^z_i \sigma^z_{i+1}  ,   
\end{equation}
where we choose periodic boundary conditions ($\sigma^z_{n+1} = \sigma^z_1 $). The two-fold degenerate ground state of the ferromagnetic Ising chain $H_{\mathrm{IM}}$, corresponding to all-spins up and all-spins down, satisfies all the Ising couplings (it is nonfrustrated) and has ground state (GS) energy $E_0=-n$.

Our objective will be to measure the expectation value of $H_{\mathrm{IM}}$ for different ratios of $A$ to $B$, which is ideally given by:
\begin{equation}
    \langle H_{\mathrm{IM}} \rangle = \frac{\Tr \left[ H_{\mathrm{IM}} e^{-\beta H_{\mathrm{TFIM}}} \right]}{\Tr \left[e^{-\beta H_{\mathrm{TFIM}}} \right]}.
\end{equation}
Earlier studies \cite{Matsuda_2009,PhysRevA.100.030303} have shown that that quantum annealers can fail to sample degenerate classical states uniformly due to the transverse-field breaking the degeneracy and the dynamics freezing for $A \ll B$. We avoid this problem by considering the thermal state associated with points before the end of the anneal, where the Hamiltonian remains quantum. 

\subsection{Obtaining mid-anneal measurements}
\label{subsec:measuring}

The D-Wave 2000Q annealing processor implements a time-dependent Hamiltonian $H(s)$ (where $s \in [0,1]$ is a  dimensionless annealing parameter), which is a linear combination of the transverse-field Hamiltonian $H_{\mathrm{TF}}$ and an Ising Hamiltonian $H_{\mathrm{p}}$---more general than the $H_{\mathrm{IM}}$ that we are implementing in our case---with a Chimera connectivity graph~\cite{Choi1,Choi2}:
%
\begin{equation} 
H(s) = A(s)H_\mathrm{TF} + B(s)H_{\mathrm{p}} \ . 
\end{equation}
The time-dependent functions $A(s)$ and $B(s)$ give the annealing schedule and are fixed by the hardware, so the device implements TFIMs with different ratios of $A/B$ along its annealing schedule.  For the DW2000Q processor we use, the minimum gap of the TFIM occurs at $s_\ast \approx 0.346$, with $A(s_\ast)/h \approx 1.05$GHz.  Further details of the processor are provided in Appendix~\ref{App:DW}. 


The rate at which the annealing schedule is traversed can be customized. Instead of having the dimensionless time parameter $s$ be simply related to the time by $s(t) = t/t_a$ (where $t_a$ is the total annealing time),  we are able to redefine $s(t)$ as a piece-wise function of $t$. This allows us to pause at specific points in the anneal, where $s$ (and thus the Hamiltonian) remains fixed for some period of time~\cite{Marshall2019, Passarelli2019}.  It also allows us to use different annealing rates $ds/dt$ for different portions of the anneal, effectively enabling us to perform approximate quenches by abruptly turning down the strength of the driver Hamiltonian from a given point mid-anneal. 

Leveraging this capability, we progress towards the thermal state associated with the Hamiltonian at $s = s_p$ by performing anneals up to $s=s_p$ and pausing at $s_p$ for a period of time thereby allowing the system to thermally relax to its steady state. We then quench as rapidly as allowed by the hardware towards $s=1$, where ideally a computational basis measurement is performed. The success of this method depends on two factors: the ability of the hardware to thermalize to the desired Gibbs state, and the annealing rate during the quench being fast enough to prevent any changes to the state.

We restrict our attention to a `forward' annealing protocol.  Here, the system is initialized in the thermal state of the Hamiltonian at $s(0) = 0$, which has very high weight on the ground state of $H(0)$, and the system is annealed from $s=0$ to some intermediate $s = s_p$ at a rate of $0 < \left( \frac{ds}{dt} \right)_i < 1 \mu \rm{s}^{-1}$.  We then pause at $s = s_p$ for some time $t_p$ and finally quench as rapidly as possible to $s=1$ at the fastest rate permitted by the hardware, $\left( \frac{ds}{dt} \right)_f = 1 \mu \rm{s}^{-1}$. A diagram of this schedule is shown in Fig.~\ref{fig:s diagram fwd}. In Appendix \ref{App:RA}, we discuss an alternative protocol using `reverse' annealing; while the details of the two protocols are different, we find no qualitative differences between the results of the two protocols.

\begin{figure}[htp]
    \centering
    \includegraphics[width=0.95\columnwidth]{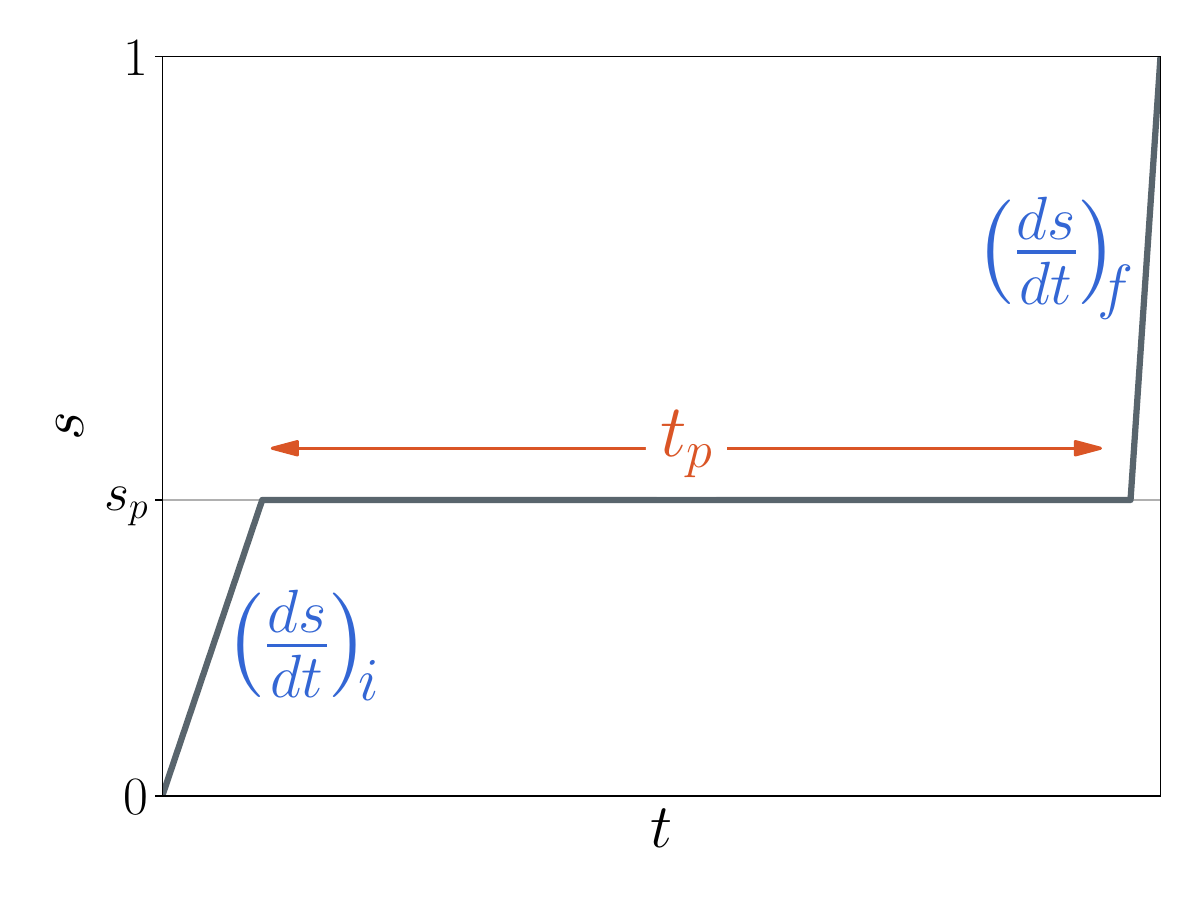}
    \caption{Diagram of the forward annealing schedule $s(t)$, showing the initial anneal to $s_p$, performed at a rate $\left( \frac{ds}{dt} \right)_i$, a pause of length $t_p$ and the final quench, performed at a rate $\left( \frac{ds}{dt} \right)_f$. The durations of the three parts are not to scale.}
    \label{fig:s diagram fwd}
\end{figure}

We incorporate gauge averaging~\cite{q-sig} to average out systematic biases that might exist on the qubits and/or couplers, such as certain qubits more readily aligning in one direction. Gauge averaging is carried out by repeating each run of $n_a$ anneals 100 times, where for each run we apply a transformation of the form $J_{ij} \rightarrow a_i a_j J_{ij}$ where $a_i \in \{-1, +1\} $ is chosen at random. This transformation corresponds to applying a unitary transformation to the Hamiltonian, so the energy spectrum is unchanged but the states are relabeled accordingly. For example, the ungauged classical state $(s_1, ..., s_n)$ is mapped to the state $(a_1 s_1, ..., a_n s_n)$.  The results in different gauges can be readily mapped back to the states of the original problem.

\section{Results}
\label{sec:results}

In what follows, we separate the discussion to two cases according to the location of the pause, which we argue exhibit qualitatively different behaviors. When the pause takes place before the minimum gap, that is, $s_p < s_\ast$, the driver Hamiltonian still dominates, and if the system is in the instantaneous GS of $H(s_p)$, the expectation value of $H_{\mathrm{IM}}$---the diagonal energy---in this region will be closer to the GS energy of $H_{\mathrm{TF}}$ than to that of $H_{\mathrm{IM}}$. On the other hand, with a pause after the minimum gap, $s_p > s_\ast$, we find the system in the region dominated by the problem Hamiltonian, and hence the expectation value of $H_{\mathrm{IM}}$ will be closer to its GS energy.

\subsection{The case of pausing before the minimum gap}
\label{subsec:beforemg}

\begin{figure}[htp]
    \centering
    \includegraphics[width=0.95\columnwidth]{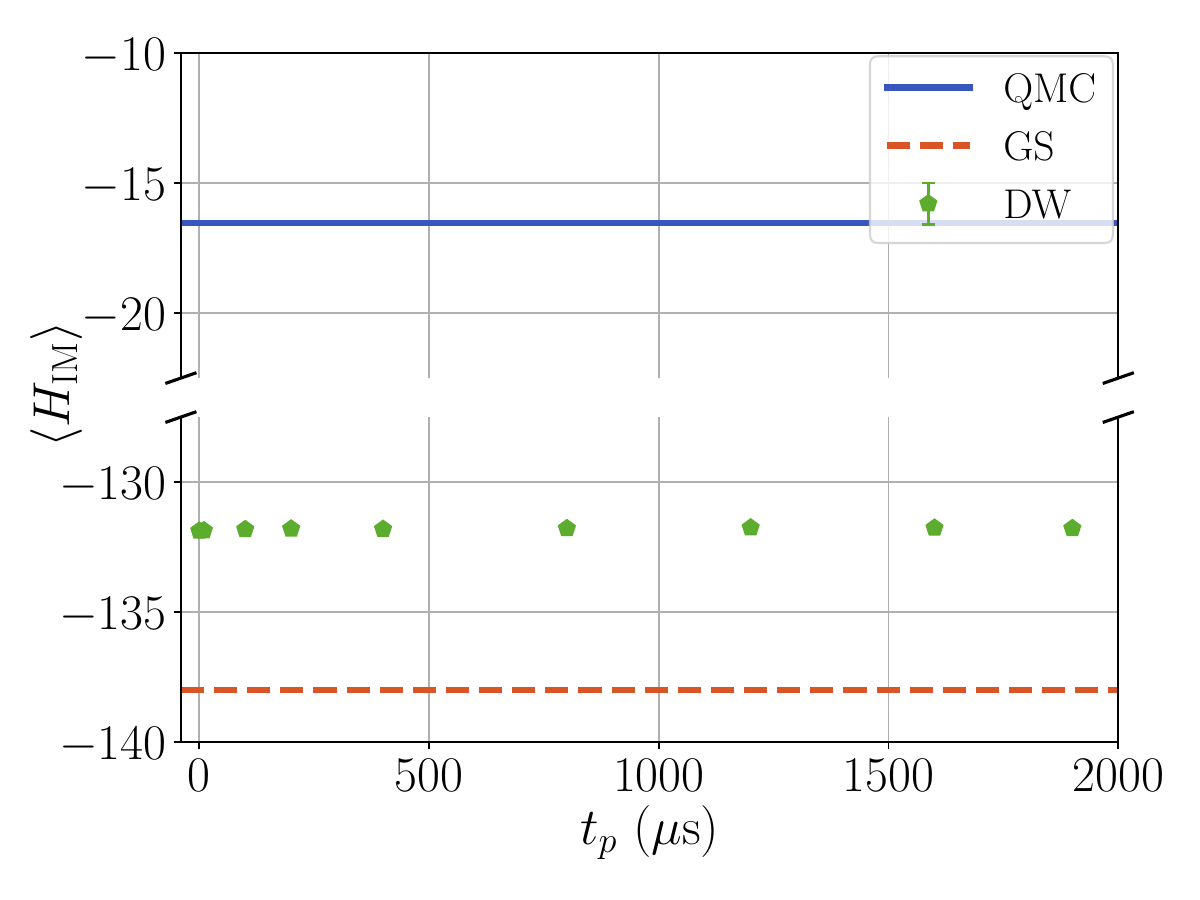}
    \par \medskip
    \includegraphics[width=0.95\columnwidth]{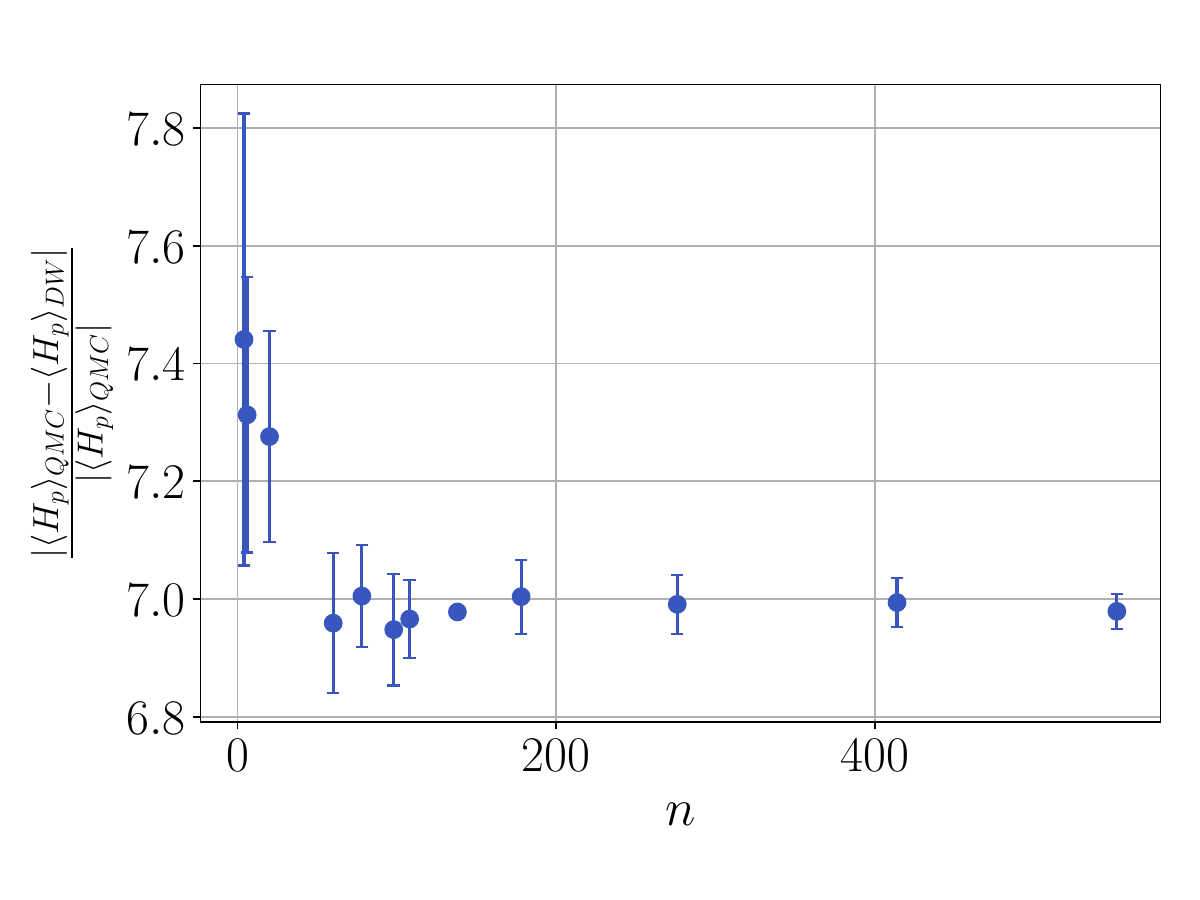}
    \caption{\textbf{Top:} Expectation value for $H_{\mathrm{IM}}$ for $n=138$ and $s_p=0.2$.  Solid blue curve corresponds to QMC prediction, dashed orange curve corresponds to the Ising GS energy, and data points correspond to the experimental results using the forward annealing protocols. \textbf{Bottom:} The relative difference between the experimental and QMC results as a function of problem size for $s_p = 0.2$, $t_p = 1.9$ms and 1500 anneals per gauge. Error bars correspond to the 95\% confidence interval calculated using a bootstrap over the 100 gauges.}
    \label{fig:before min gap}
\end{figure}

We first consider the case of $s_p < s_\ast$, i.e., where the pause takes place prior to the system reaching its minimum gap. In this region, the strength of $A(s_p) H_{\mathrm{TF}}$  is considerably greater than that of $B(s_p) H_{\mathrm{IM}}$, so there is little overlap between the GS of $H_{\mathrm{IM}}$ and the instantaneous GS of $H(s_p)$. The QMC results reflect this. For instance, with pause location $s_p=0.2$ and problem size $n=138$, the expectation value of $H_{\mathrm{IM}}$ is $\langle H_{\mathrm{IM}} \rangle_{\text{exact}} \big|_{s_p=0.2} \approx -16.5$, while the Ising GS energy is $-138$. When we calculate this same expectation value using the output from the annealer, however, we find $\langle H_{\mathrm{IM}} \rangle_{DW} \big|_{s_p=0.2} \approx -132$, suggesting that the experimental measurement is failing to capture the state of the system in the region before the minimum gap, with the relative difference between experimental and QMC results being $700\%$. This is depicted in Fig.~\ref{fig:before min gap} (top).

This experimental behavior of being very far from the expected value and very close to the Ising ground state value holds for all the problem sizes we studied, ranging from 4 to over 500 qubits. Figure~\ref{fig:before min gap} (bottom) shows the relative difference between experiment and theory as a function of size. While very small problems seem to do slightly worse, this difference remains fairly constant for a large range of sizes. We note that the experimental results do not change in any significant way as we increase the pause time $t_p$, so we deduce that the system is very close to its steady state at the pause point.

A possible explanation for this dramatic difference is the upper limit on the annealing rate, which constrains our quench to a maximum $\left( \frac{ds}{dt} \right)_f = 1 \mu \rm{s}^{-1}$. This rate is likely not sufficiently fast to prevent the state from evolving during the quench and allowing the dynamics to populate the Ising GS. We validate this conjecture using master equation simulations in Sec.~\ref{subsec:quench}.

\subsection{The case of pausing after the minimum gap}
\label{subsec:aftermg}

\begin{figure}[htp]
  \centering
  \includegraphics[width=.95\columnwidth]{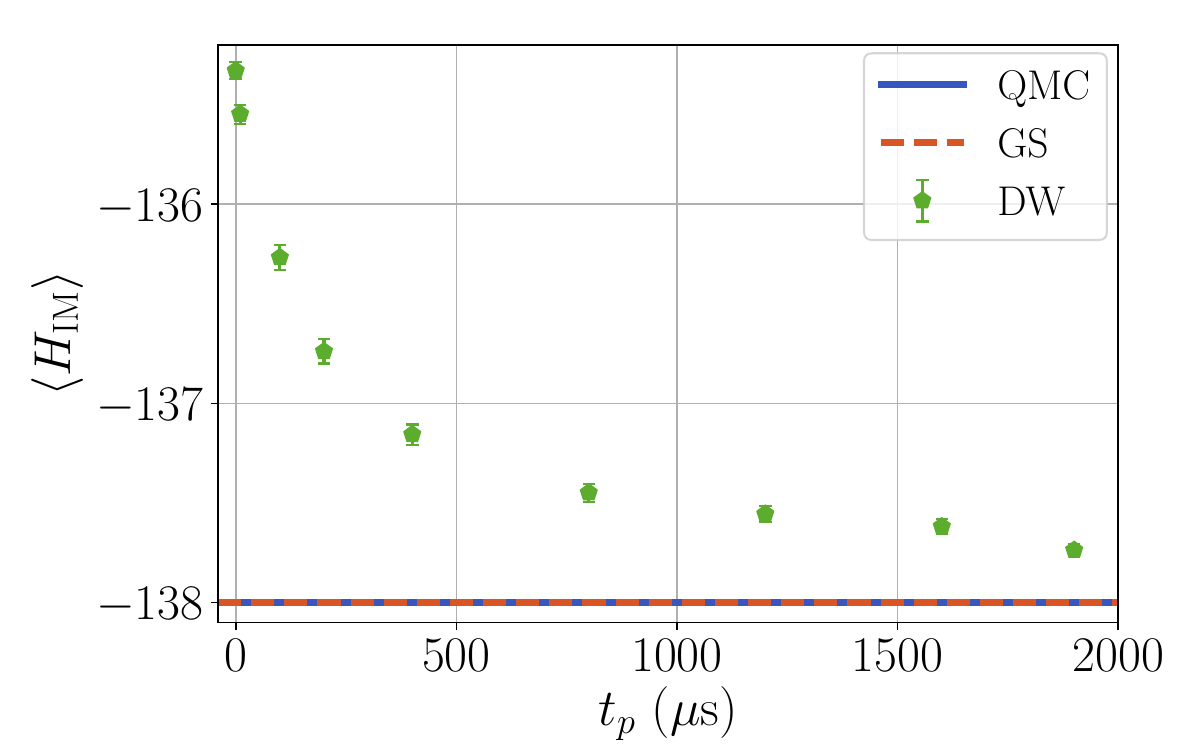}  
  \newline
  \centering
  \includegraphics[width=.95\columnwidth]{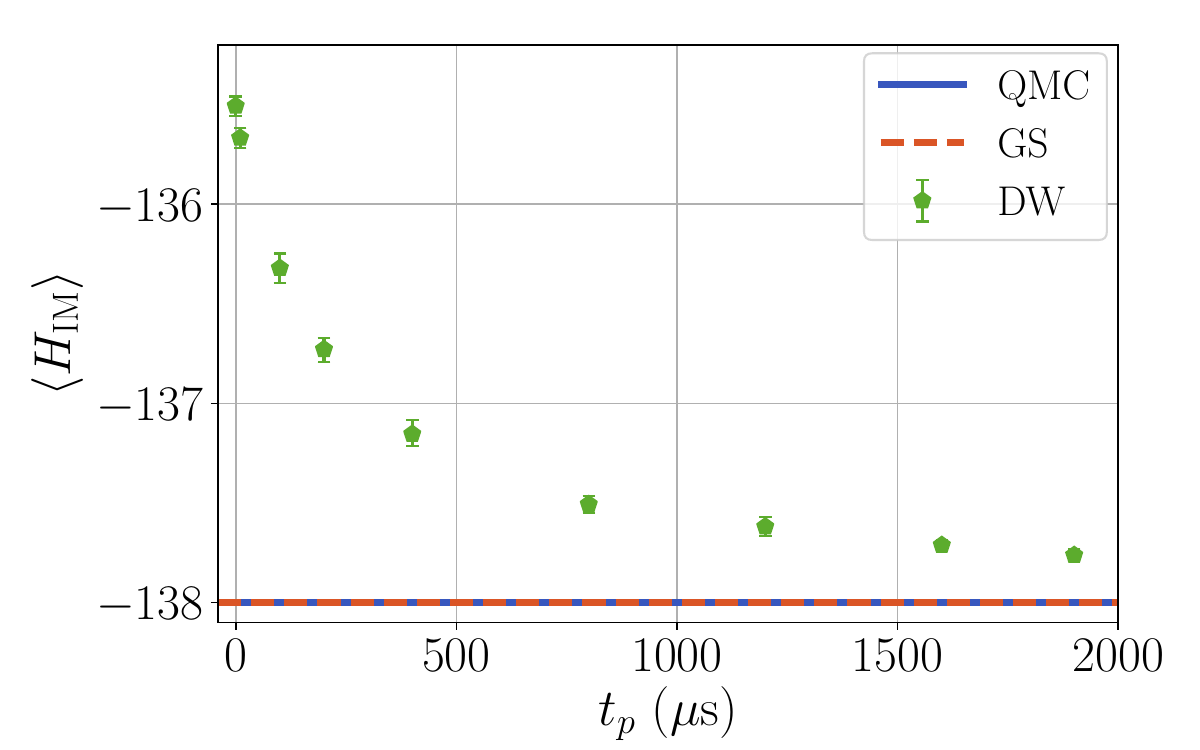}  
  \newline
  \centering
  \includegraphics[width=.95\columnwidth]{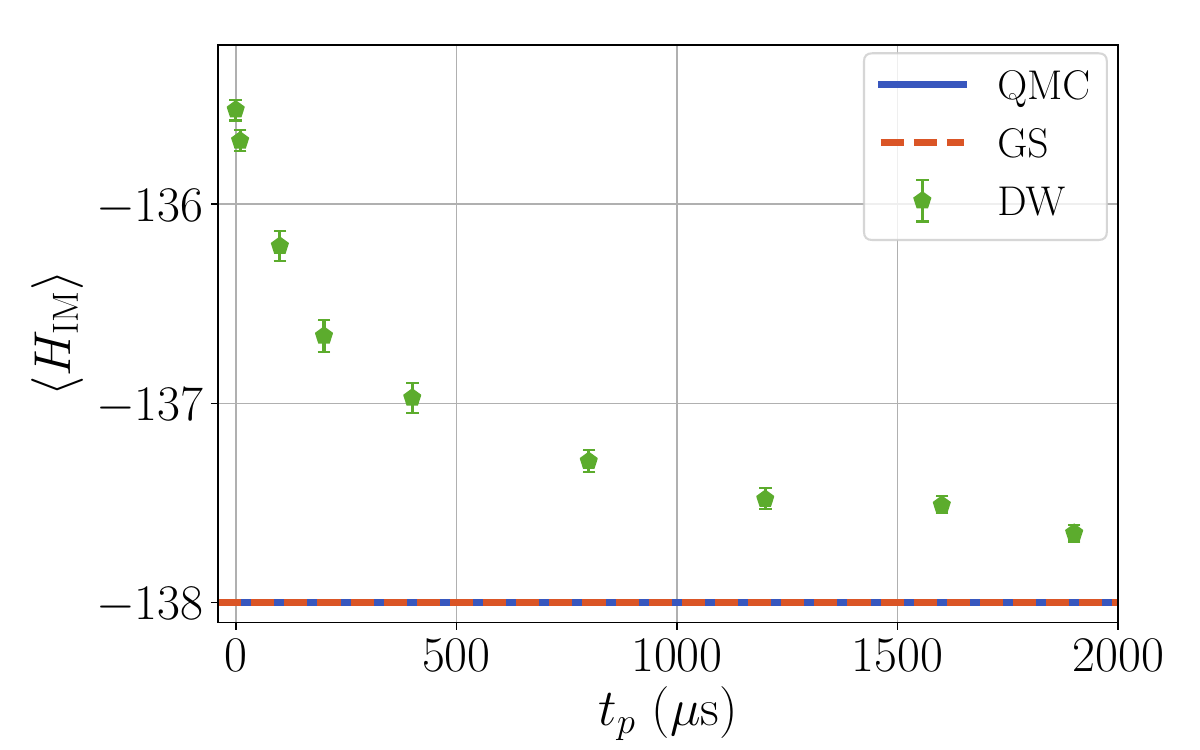}  
  \newline
  \centering
  \includegraphics[width=.95\columnwidth]{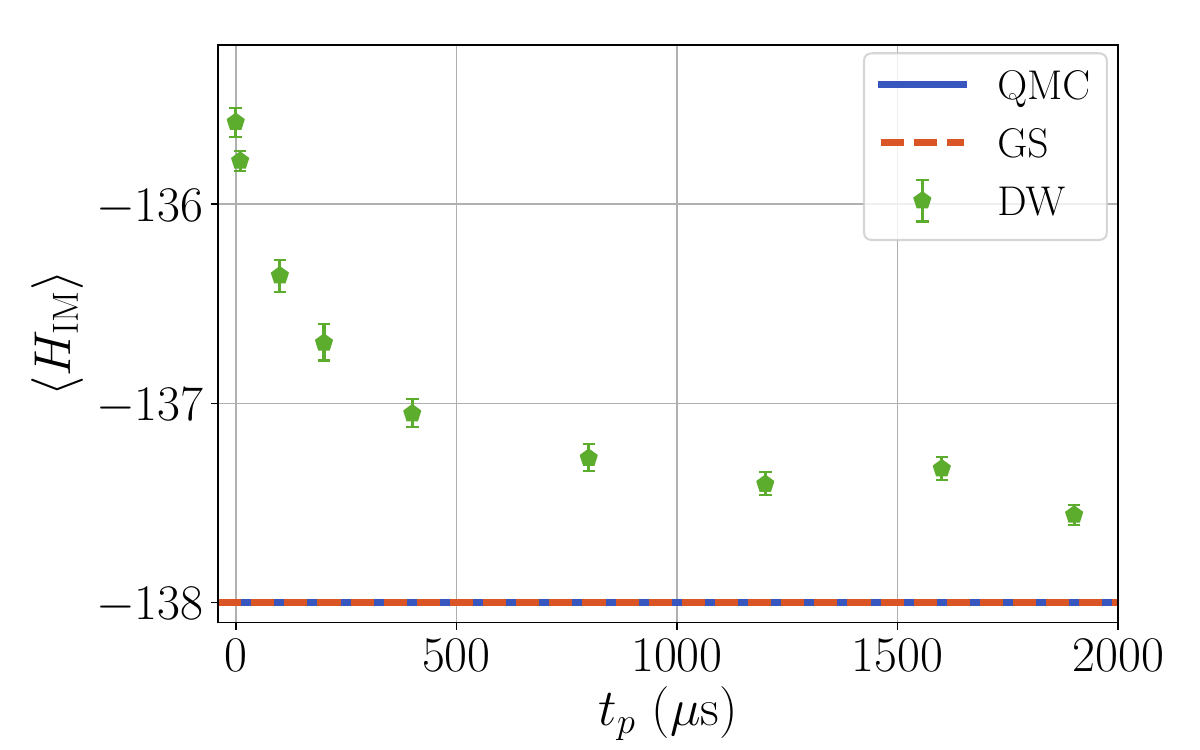}
  \newline
\caption{From top to bottom, $\langle H_{\mathrm{IM}} \rangle$ comparison between experiment and simulations for $n=138$ and $s_p=0.5, 0.6, 0.7$ and $0.8$. Solid blue curve corresponds to QMC result, dashed orange curve corresponds to the Ising GS energy, and data points correspond to the DW results using the  forward  annealing protocols.  Error bars correspond to the 95\% confidence interval calculated using a bootstrap over the 100 gauges.}
\label{fig:after min gap}
\end{figure}

When we choose the pause to occur after the minimum gap, i.e., in the case where $s_p > s_\ast$, the  Hamiltonian $B(s_p) H_{\mathrm{Ising}}$ dominates over $A(s_p) H_{\mathrm{TF}}$.  We therefore expect $\langle H_{\mathrm{IM}} \rangle$ to be significantly closer to $E_0= -n$. The experimental results in this region turn out to depend sensitively on the value of $s_p$, as we show in Fig.~\ref{fig:after min gap}. We observe the best agreement with QMC around $s_p=0.6$. When we pause at later points, however, $\langle H_{\mathrm{IM}} \rangle$ obtained from the annealer moves away from the correct value.

The dependence on $s_p$ is likely again due to the quench rate.  For $s_p < 0.6$, the thermal state at $s_p$ still does not have a complete overlap with the Ising ground state, so the quench from $s= s_p$ to $s= 1$ is sufficiently slow for the system to repopulate the Ising ground state.  Hence we find that the experimental $\langle H_{\mathrm{IM}} \rangle$ is lower than predicted by QMC.  For $s_p > 0.6$, the Hamiltonian $A(s_p) H_{\mathrm{TF}}$ is very weak relative to $B(s_p) H_{\mathrm{IM}}$, so the dynamics are expected to be extremely slow and effectively frozen~\cite{Amin:2015qf,Marshall2019}.  In this case, the system likely does not have enough time to thermalize.  The points around $s_p = 0.6$ represent a `sweet spot' where the system is still able to thermalize and is only minimally affected by the slowness of the quench.

\subsection{Quantifying the impact of the quench}
\label{subsec:quench}

In the previous section we have seen how the experimental samples fail to reproduce the correct thermal expectation values associated with the point where we pause $s_p$. 
We argued that this is likely due to the quench rate from $s=s_p$ to $s=1$ not being sufficiently fast, with the system continuing to evolve during the quench.
In order to confirm this hypothesis, we use the adiabatic master equation (ME)~\cite{Albash2012} to simulate the quantum annealing process using different quench rates. The key feature of this model is that for any fixed $s$ Hamiltonian, the fixed point of the dissipative dynamics is the Gibbs state of $H(s)$.  Therefore, for any sufficiently long pause and sufficiently fast quench, we expect the simulation results to agree with the theoretical prediction.
Because of the computational cost of the simulations, we are only able to obtain exact results for the smallest systems with $n=4$, but already at these sizes we are able to see the adverse effects of the quench rate for reproducing the correct thermal expectation values.  We provide further details of our parameter choices for the simulations in Appendix \ref{App:ME}.

\begin{figure}[htp]
    \centering
    \includegraphics[width=0.95\columnwidth]{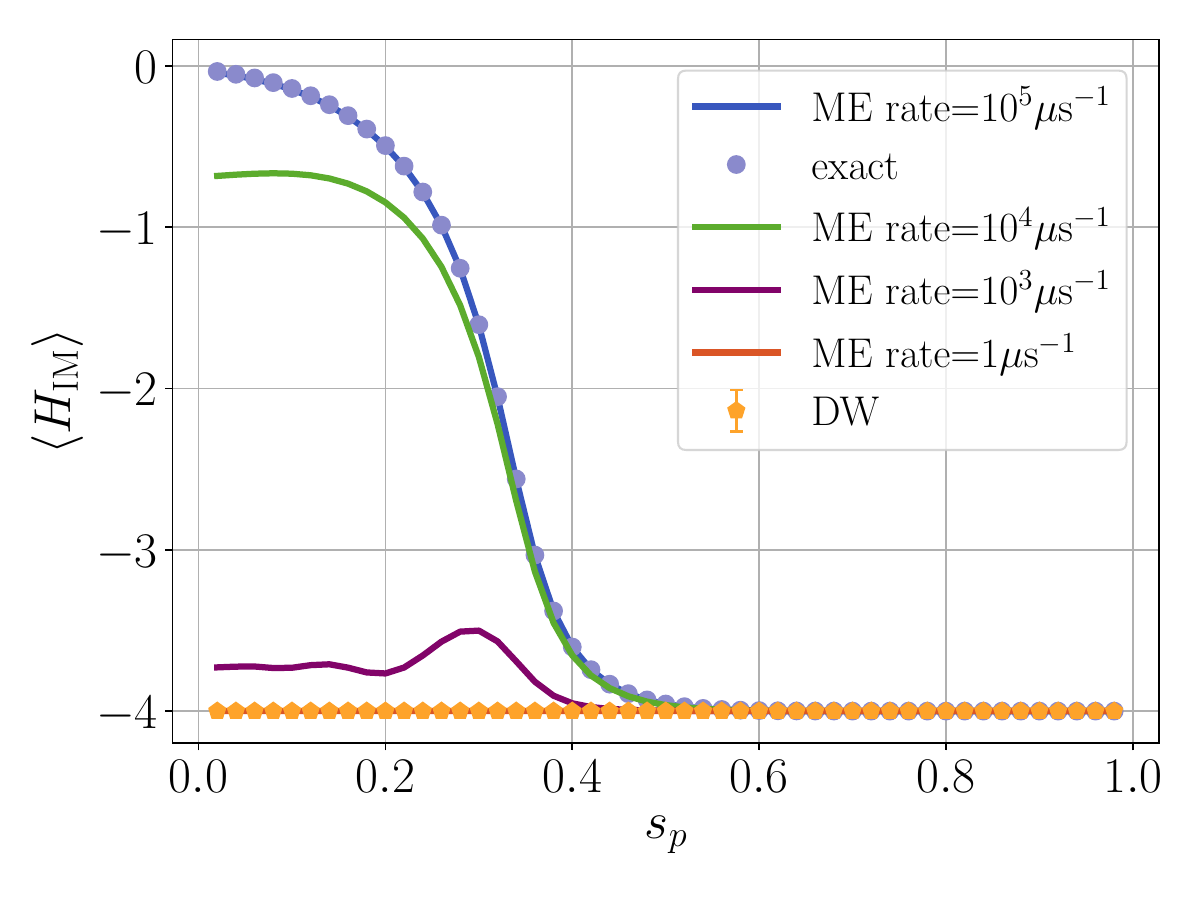}
    \par \medskip
    \caption{Comparison of numerical (ME) simulations using several different quench rates. The slowest one, $1 \mu s^{-1}$, matches that of the experiments, while the fastest one, $10^5 \mu s^{-1}$, agrees with the analytical results. Simulations are for a system size of $n=4$.  Also shown are the exact results and the experimental ones.  Simulation parameters are given in Appendix \ref{App:ME}.}
    \label{fig:master eq}
\end{figure}

We first run these simulations using the same schedule as the annealer, with a quench annealing rate of \hbox{$\frac{ds}{dt} = 1 \mu \rm{s}^{-1}$}. These are shown as the `slow quench' results in Fig.~\ref{fig:master eq}, where we find the simulated $\langle H_{\mathrm{IM}} \rangle$ results are in agreement with experiment at all pause times $s_p$ during the anneal.  Specifically, we find $\langle H_{\mathrm{IM}} \rangle$ to be close to the Ising GS energy regardless of the value of $s_p$.  In fact, for this small system size, even if we considered closed system dynamics, i.e. we decoupled the system from the thermal environment,  we would get the same result, pointing to the fact that the quench is so slow that the evolution across the minimum gap is effectively adiabatic.

Next we survey a wide range of quench rates to find how fast the quench must be to reproduce the correct thermal expectation values.  For the $n=4$ system, we find that we need to go up to $\frac{ds}{dt} = 10^3 \mu \rm{s}^{-1}$ before seeing any change in the behavior of $\langle H_{\mathrm{IM}} \rangle$. As the annealing rate $\frac{ds}{dt}$ becomes larger, $\langle H_{\mathrm{IM}} \rangle$ gets closer to the exact results, and they finally become in good agreement when the annealing rate is $\frac{ds}{dt} = 10^5 \mu \rm{s}^{-1}$ (the `fast quench' in Fig.~\ref{fig:master eq}). This  implies that we need to  increase the current fastest rate allowed by the hardware by a factor of $10^5$, with the minimum time for a full anneal decreasing from $1 \mu \rm{s}$ to around 10~ps.

While performing simulations at larger system sizes becomes computationally prohibitive, we can provide a simple argument that suggests that the annealing rate must be even faster at larger system sizes.  For simplicity, let us assume a constant quench rate and an evolution from $s=s_p$ to $s=1$ that is purely unitary. In order for the unitary dynamics to not change the state significantly we must require
\beq \label{eqt:norm}
\left| \mathrm{T} \exp \left[ -i \left(\frac{ds}{dt}\right)^{-1} \int_{s_p}^1 H(s') ds' \right] - \mathds{1} \right| \leq  \epsilon
\eeq
for some suitably small $\epsilon$, which may also need to decrease with system size in order to reproduce the thermal expectation values to the desired accuracy.  Here $| \cdot |$ denotes the operator norm, but any appropriate distance measure can be used.  If we expand the time-ordered exponential for small $(d s / dt )^{-1}$, it follows from the extensivity of the Hamiltonian that to ensure each term inside the norm  remains small, the inverse of the annealing rate $(d s / dt )^{-1}$ must scale at least as $1/n$. Therefore, we already find that simply to ensure that the unitary dynamics does not change the state significantly, the quench rate must become faster as the system size is increased.

\subsection{Other observables: Magnetization}
\label{subsec:mag}

We have so far considered the thermal expectation value of $H_{IM}$ as a convenient benchmark for DW's behavior as a thermal sampler. The question arises whether other observables of the TFIM are more robust to the control limitations we have highlighted.  For example, in Refs.~\cite{King2018,Harris2018} certain observables showed good agreement, while others did not. This is an important point to consider when evaluating DW's potential as a quantum thermal sampler versus a quantum simulator of certain observables; for a fully functional sampler, consistent behavior across different observables would be required.

\begin{figure}[htp]
    \centering
    \includegraphics[width=0.95\columnwidth]{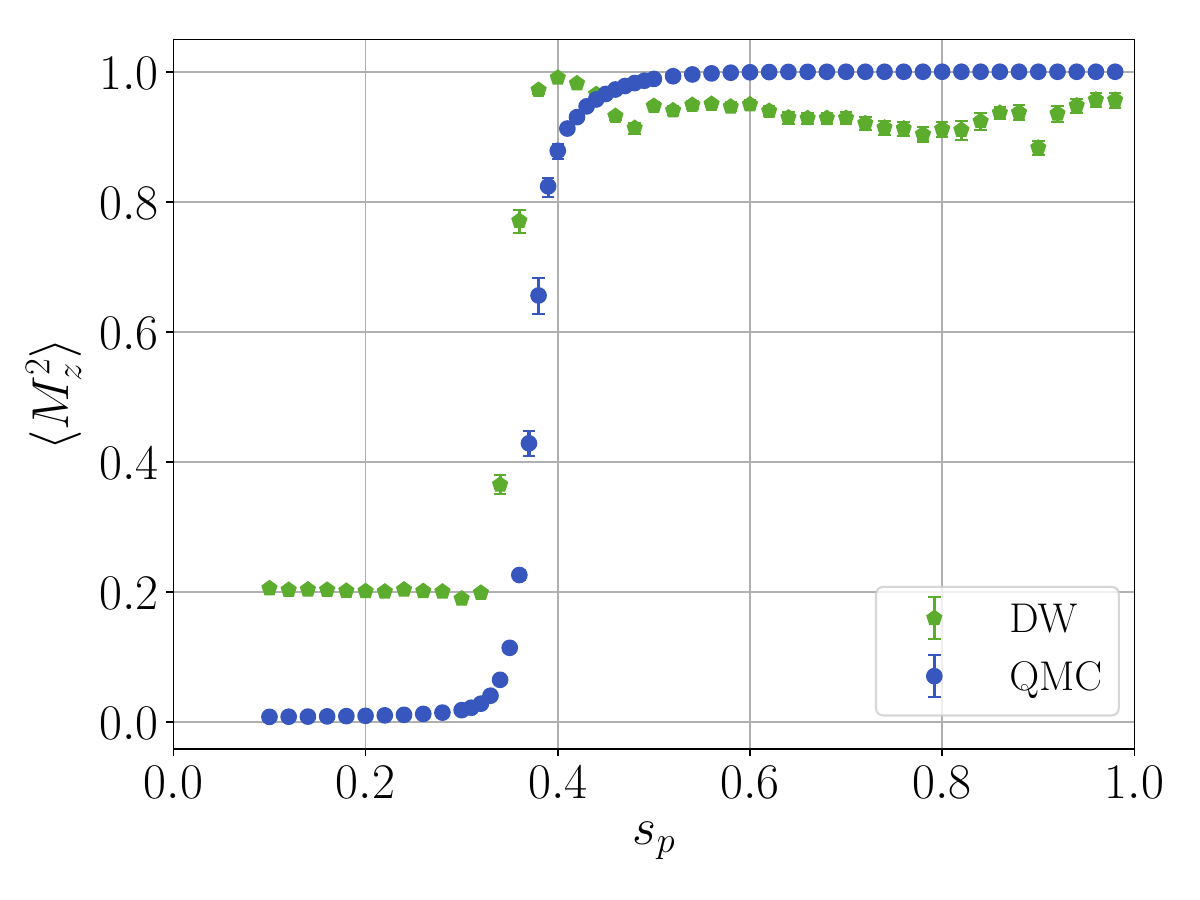}
    \par \medskip
    \includegraphics[width=0.95\columnwidth]{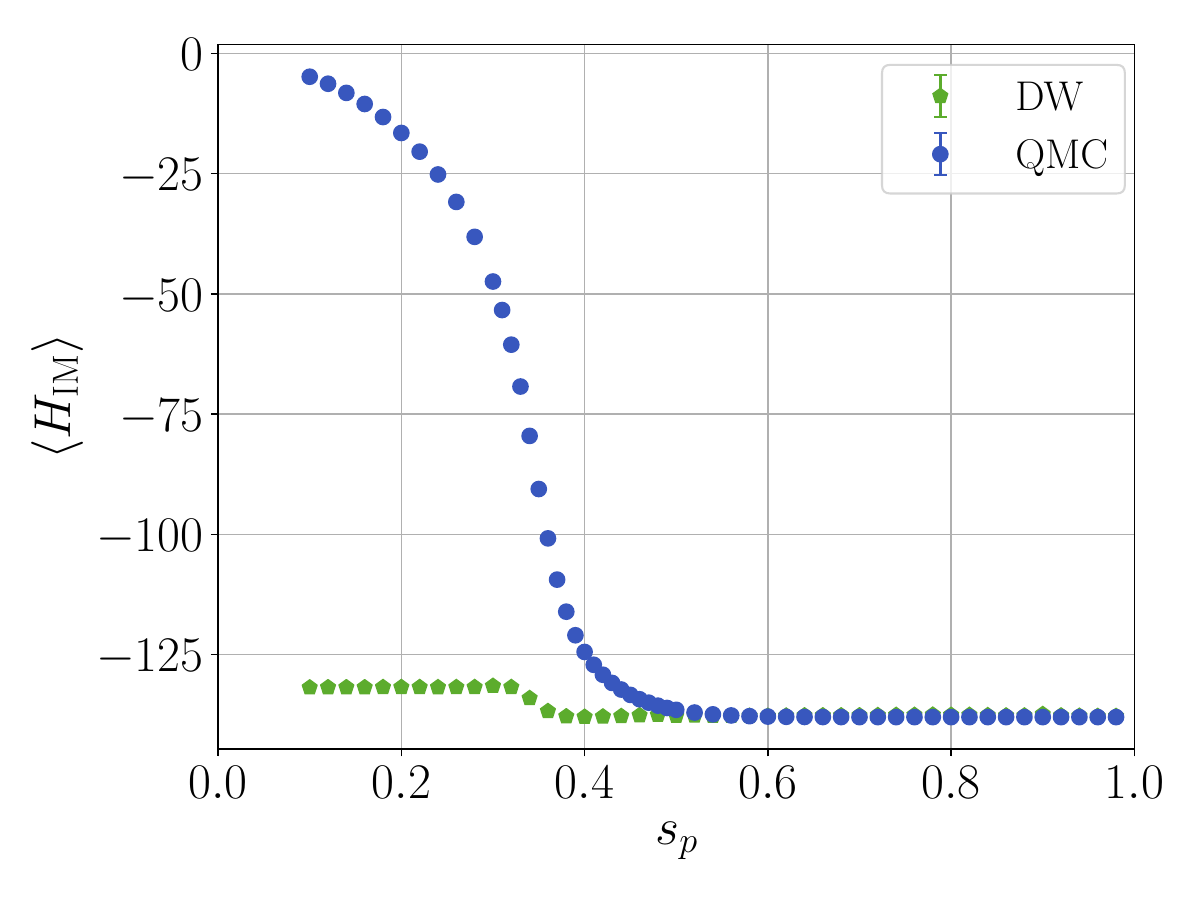}
    \caption{Comparison of DW and QMC results for an $n=138$ chain at a range of locations through the anneal, for $\langle M_z^2 \rangle$ (\textbf{top}) and $\langle H_{IM} \rangle$ (\textbf{bottom}). Blue data points correspond to QMC results and green data points to DW with $t_p = 1.9$ms and 1500 anneals per gauge. Error bars correspond to the 95\% bootstrap confidence interval.}
    \label{fig:mag}
\end{figure}

We choose the squared longitudinal magnetization $M_z^2$---due to the symmetry of the system, $\langle M_z \rangle$ is always 0. This observable can be taken to be the relevant order parameter for the TFIM. Looking at the overall picture (Fig.~\ref{fig:mag}), it is apparent that the two observables follow different patterns of behavior when DW's results are compared to QMC; while $\langle H_{IM} \rangle$ always stays close to the value it should attain at the end of the anneal, $\langle M_z^2 \rangle$ follows a trend that is qualitatively more similar to the QMC data, suggesting that $M_z^2$ is less sensitive than $H_{\mathrm{IM}}$ to distortions in the sampled distribution.

\begin{figure}[htp]
    \centering
    \includegraphics[width=0.95\columnwidth]{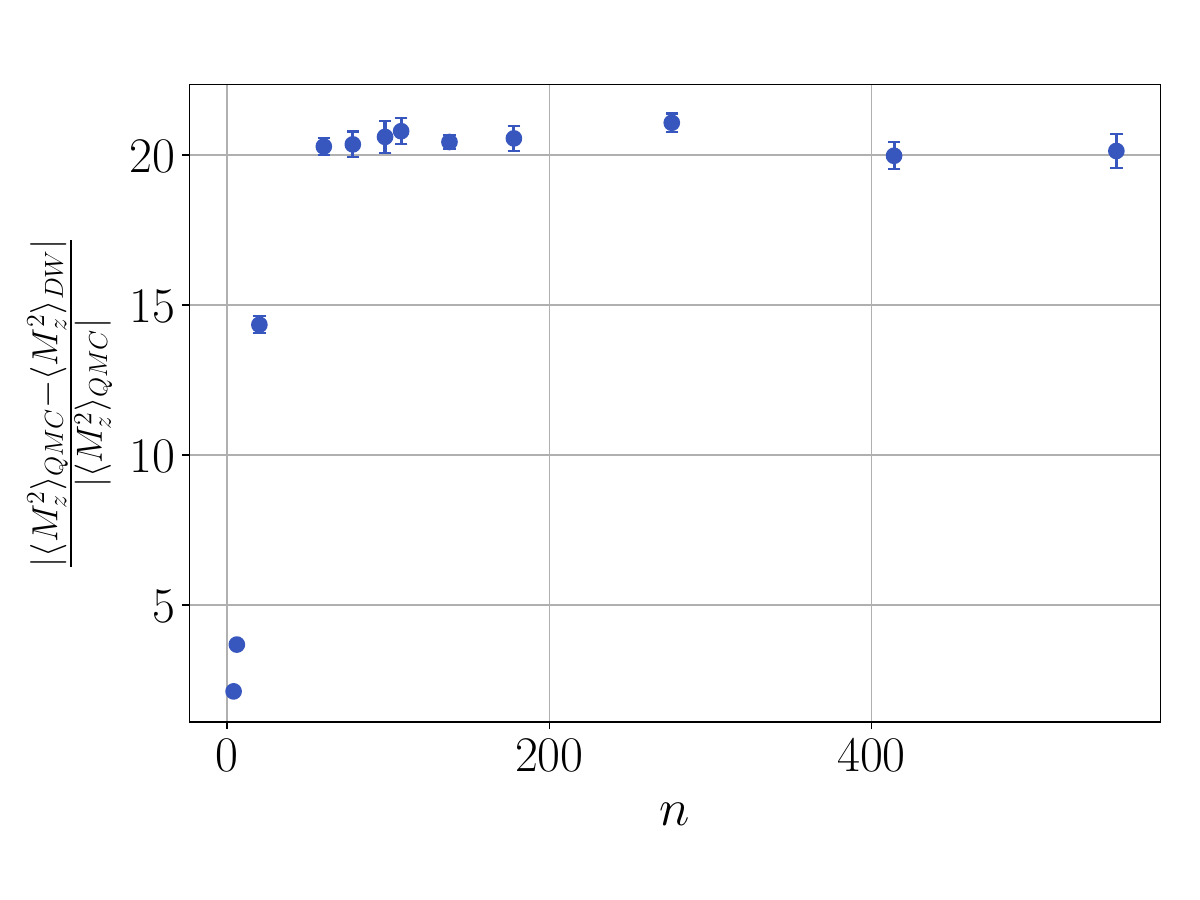}
    \caption{Relative difference between the experimental and QMC results for $\langle M_z^2 \rangle$ as a function of problem size for $s_p = 0.2$, $t_p = 1.9$ms and 1500 anneals per gauge. Error bars correspond to the 95\% bootstrap confidence interval.}
    \label{fig:mag_diff}
\end{figure}

However, the $\langle M_z^2 \rangle$ results produced by DW do not match QMC anywhere in the anneal (except at one point where they cross), and in fact their relative difference is much larger than for $\langle H_{\mathrm{IM}} \rangle$ for most system sizes (Fig.~\ref{fig:mag_diff}).  Furthermore, the the transition from small   $\langle M_z^2 \rangle$ to  large $\langle M_z^2 \rangle$ is shifted, indicating that the location of the finite-size precursor of the phase transition is distorted.

\section{Conclusion}
\label{sec:conclusion}
In this work, we asked whether the current ability to control the annealing schedule of the D-Wave 2000Q quantum annealing processor allows us to accurately probe the state of the system in the middle of the anneal.  The argument put forth is that if the system is quenched sufficiently rapidly from $s = s_p$ towards $s=1$ where the measurement is performed, we would be effectively performing a measurement on the state at $s_p$. We used the nonfrustrated one-dimensional transverse-field Ising model as a case study and found that the D-Wave 2000Q quantum annealing processor cannot be used to reliably generate samples that reproduce the correct thermal expectation values for the diagonal observables $H_{\mathrm{IM}}$ and $M_z^2$.  We show in Appendix~\ref{App:FC} that our findings remain unchanged when we consider frustrated chains.

We identified the most likely culprit to be the quench rate: with a maximum annealing rate of $\frac{ds}{dt} = 1 \mu \rm{s}^{-1}$ offered by the device, the system continues to evolve before the measurement occurs. We also verify in Appendix~\ref{App:ET} that fluctuations in the temperature cannot explain the discrepancy between the QMC predictions and our experimental observations. 

The different behavior observed for  $H_{\mathrm{IM}}$ and $M_z^2$ confirms our suspicion that the degree of agreement between DW and QMC can vary across different observables. This adds another layer of complication if we wish to use the annealer to predict thermal expectation values, as its behavior regarding a particular observable is not indicative of what would happen for a different one, and each case would need to be considered individually. Our results, considered along those of Refs.~\cite{King2018,King2019,Harris2018}---where experiments on the same platform were able to reproduce the correct thermal expectation values of certain observables (but not others) in far more complicated systems---suggest that the ability of the annealer to produce correct mid-anneal predictions is highly dependent on the observable and the problem considered, and their particular susceptibility to the quench.  For the Hamiltonian systems in Refs.~\cite{King2018,King2019,Harris2018}, the quench did not change the samples in a significant way for the expectation values of observables that were calculated.  However, the 1D TFIM does not exhibit this robustness to the observables we study, and this can likely be attributed to the ease with which domain wall excitations can be formed and moved across the system. Even beyond the issues related to the quench, a recent study~\cite{Bando2020} showed that the distribution of kinks generated by DW during a standard anneal deviates from those expected for a spin-boson open quantum system model, in addition to varying between different D-Wave processors, shows that other sources of noise can further distort the distribution \cite{2020arXiv200607685C,2020arXiv201208827V}.  This at least shows that simply the convergence of expectation values may not necessarily mean convergence to an accurate expectation value. 


Beyond the experimental difficulty of achieving the necessary fast quench rates, another fundamental problem arises.  In our work, we have assumed an ideal qubit (2-level) Hamiltonian, but in the hardware the effective qubit Hamiltonian is realized by projecting onto the lowest two energy levels of a superconducting flux qubit~\cite{Boixo:2014yu}.  In this effective description, the computational basis is defined in terms of symmetric and anti-symmetric combinations of the ground and first-excited states of the flux qubit Hamiltonian at zero flux, and this basis changes along the annealing schedule~\cite{Boixo:2014yu,Vinci:2017aa}.  In the unitary dynamics, the above approximation manifests itself as geometric terms in the effective qubit Hamiltonian that contributes to the evolution even in the limit of a sudden quench~\cite{Vinci:2017aa}.  These effects are not captured by our ideal qubit assumption in Eq.~\eqref{eqt:norm}, and while the effect on a single qubit may be small, the error associated with it accumulates with system size.

An additional difficulty will be added as we seek to study problems of increasing complexity. When embedding is required, its negative effect on the likelihood of obtaining representative samples is particularly harmful for sampling problems~\cite{Marshall2019_2}, where we are interested in states at all energies as opposed to only ground states like for optimization problems. This effect is worsened with problem size and the complexity of the embedding.

At present, there is no formal identification of which observables for which systems will be susceptible to the slowness of the quench and distortions of the thermal statistics, and hence extreme care must be taken when interpreting results from the variable annealing schedule as `measurements-in-the-middle.'  While we conclude that currently available quantum annealing devices are not well posed to function as thermal samplers, our results do not preclude the possibility that they can still serve as useful simulators to study properties that are more robust to these imperfections, such as those with time scales much longer than the quench rate. It remains an important research direction to identify such accessible quantities in systems that cannot be tackled via classical simulation in order for such quantum simulators to achieve their promise.


\begin{acknowledgements}
We thank Paul Warburton and Jeffrey Marshall for useful comments on the manuscript. TA thanks Daniel Lidar for useful discussions. ZGI would like to acknowledge support from the USRA Feynman Quantum Academy at NASA Ames Research Center.
The research is based upon work (partially) supported by the Office of
the Director of National Intelligence (ODNI), Intelligence Advanced
Research Projects Activity (IARPA), via the U.S. Army Research Office
contract W911NF-17-C-0050. This material is based on research sponsored by the Air Force Research laboratory under
agreement number FA8750-18-1-0044. The views and conclusions contained herein are
those of the authors and should not be interpreted as necessarily
representing the official policies or endorsements, either expressed or
implied, of the ODNI, IARPA, or the U.S. Government. The U.S. Government
is authorized to reproduce and distribute reprints for Governmental
purposes notwithstanding any copyright annotation thereon.
\end{acknowledgements}

%

\appendix
\section{The D-Wave 2000Q Quantum Annealing Processor}
\label{App:DW}

Our experimental results are generated by a D-Wave 2000Q Quantum Annealing Processor (DW), located at NASA Ames Research Center. Quantum annealing~\cite{finnila_quantum_1994,Brooke1999,kadowaki_quantum_1998,Farhi:01,Santoro} is a metaheuristic for solving certain types of discrete optimization problems. Drawing inspiration from simulated annealing, it uses quantum fluctuations to encourage exploration of the solution space of an Ising-type problem Hamiltonian, 
\begin{equation}
H_{\mathrm{IM}} = \sum_{i, j} J_{i, j} \sigma^z_i \sigma^z_j + \sum_i h_i \sigma_i^z.    
\end{equation}
When the solution to a problem of interest can be encoded as the ground state (GS) of $H_{\mathrm{IM}}$ and initializing the system in the known GS of a different Hamiltonian, we can make use of the adiabatic theorem~\cite{Jansen:07} to provide a guarantee that the GS can be reached with high probability for a sufficiently slow interpolation.

\begin{figure}[htp]
    \centering
    \includegraphics[width=0.95\columnwidth]{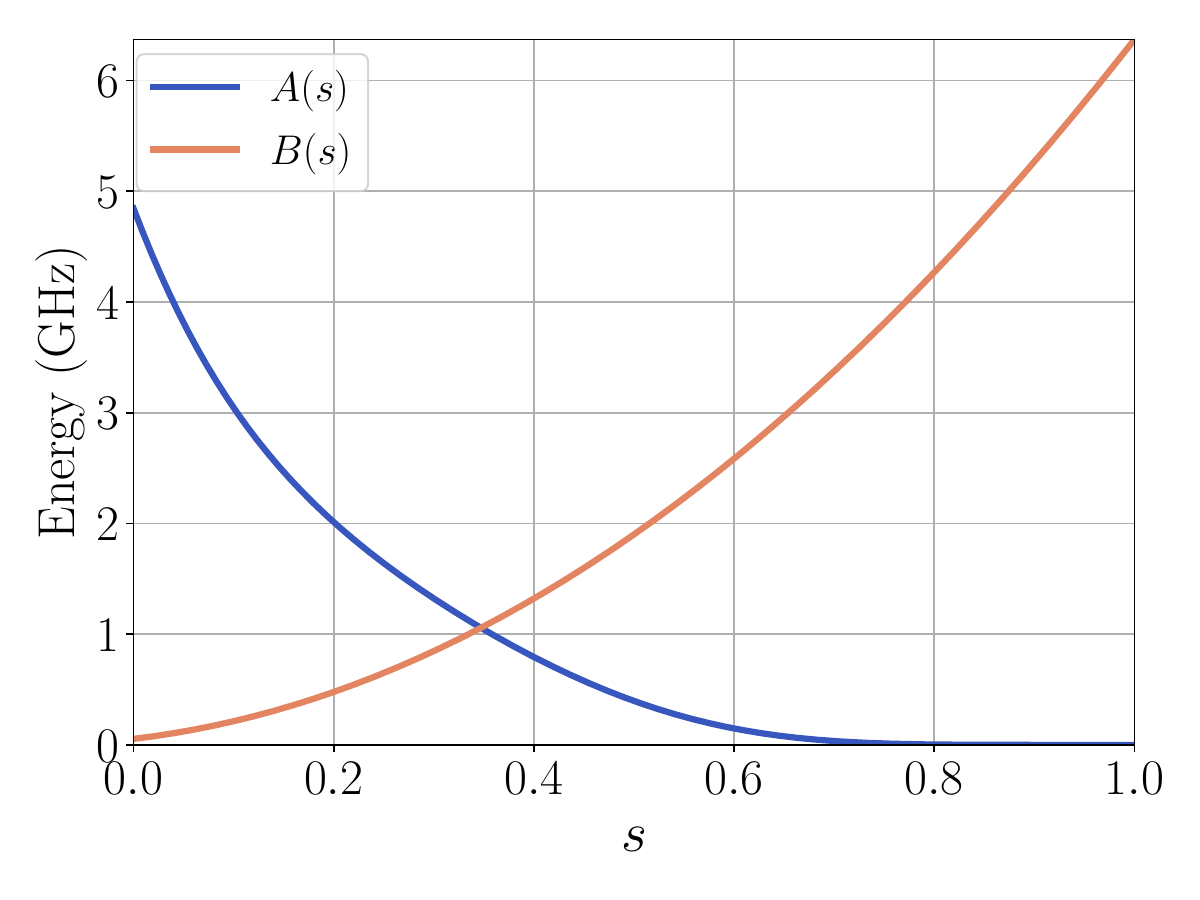}
    \caption{Strengths of the driver and problem Hamiltonians as functions of the dimensionless time parameter $s$, shown in units of $h = 1$.}
    \label{fig:schedules}
\end{figure}

In the case of our annealer, the system is initialized in the GS of a transverse-field driver Hamiltonian $H_d =-\sum_{i=1}^n \sigma_i^x$, and evolved through time by decreasing the strength of $H_d$ while increasing that of $H_{\mathrm{IM}}$. The time-dependent Hamiltonian is:
\begin{equation}
H(s) = A(s)H_d + B(s)H_{\mathrm{IM}},  
\end{equation}
where $s=t/t_a$ is a dimensionless time parameter and $t_a$ is the total annealing time. $A(s)$ and $B(s)$ determine the respective strengths of the driver and problem Hamiltonian (Fig.~\ref{fig:schedules}), with $A(0) \gg B(0)$ and $A(1) \ll B(1)$.

\begin{figure}[htp]
    \centering
    \includegraphics[width=0.95\columnwidth]{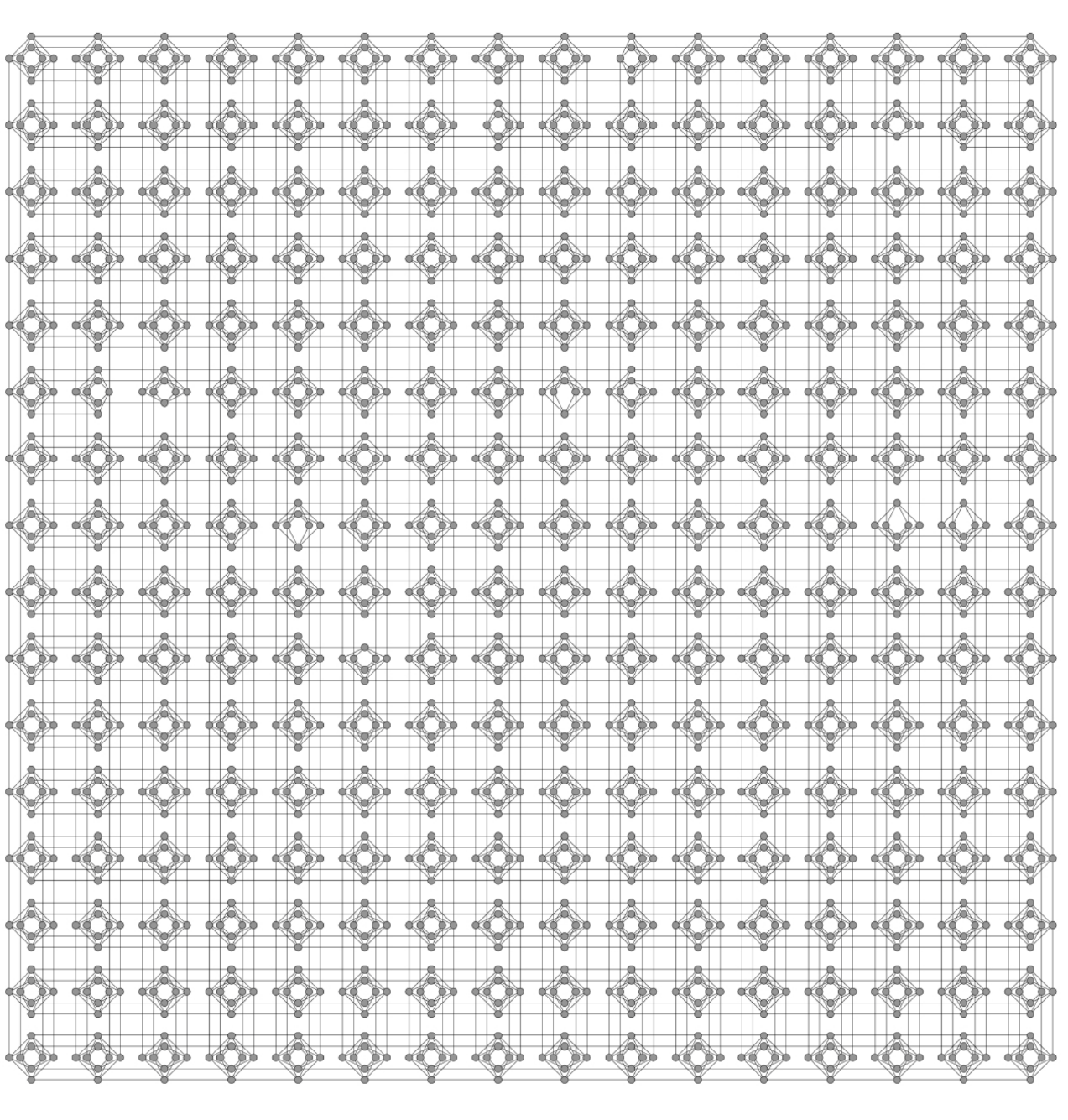}
    \caption{The quantum annealer hardware adjacency graph, a $16\times 16$ Chimera graph. The nonworking qubits are not shown here.}
    \label{fig:chimera}
\end{figure}

The D-Wave 2000Q processor consists of a lattice of niobium superconducting quantum interference device (SQUID) qubits operating at a temperature of approximately \hbox{$12$mK}, and connected according to a Chimera architecture~\cite{Harris2010, Bunyk2014}. Chimera graphs are made up of smaller unit cells, which can be repeated in two dimensions to achieve graphs of different sizes. Each cell is a complete bipartite graph $K_{4,4}$, where a qubit is connected to four others within its cell and two more in adjacent cells. The Chimera graph $C_L$ is obtained by arranging the unit cells in a square pattern with $L$ cells per side. The processor we are using features a $C_{16}$, with a total of 256 cells and 2048 qubits (although a few are inoperative due to fabrication issues). Its complete connectivity graph is shown in Fig.~\ref{fig:chimera}.

\section{Reverse Annealing Protocol} 
\label{App:RA}

\begin{figure}
    \centering
    \includegraphics[width=0.95\columnwidth]{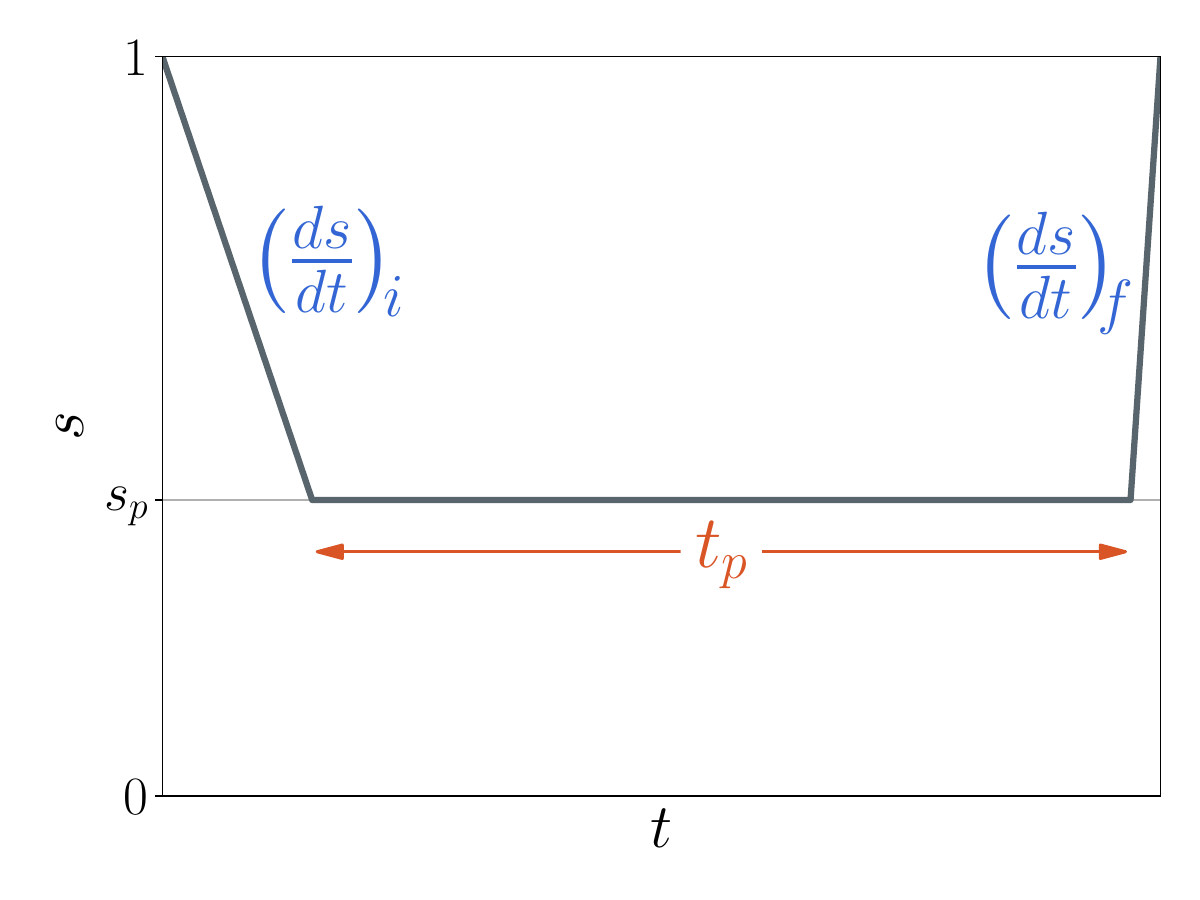}
    \caption{Diagram of the reverse annealing schedule $s(t)$, showing the initial anneal to $s_p$, performed at a rate $\left( \frac{ds}{dt} \right)_i$, a pause of length $t_p$ and the final quench, performed at a rate $\left( \frac{ds}{dt} \right)_f$. The durations of the three parts are not to scale.}
    \label{fig:s diagram rev}
\end{figure}

\begin{figure}
    \centering
    \includegraphics[width=0.95\columnwidth]{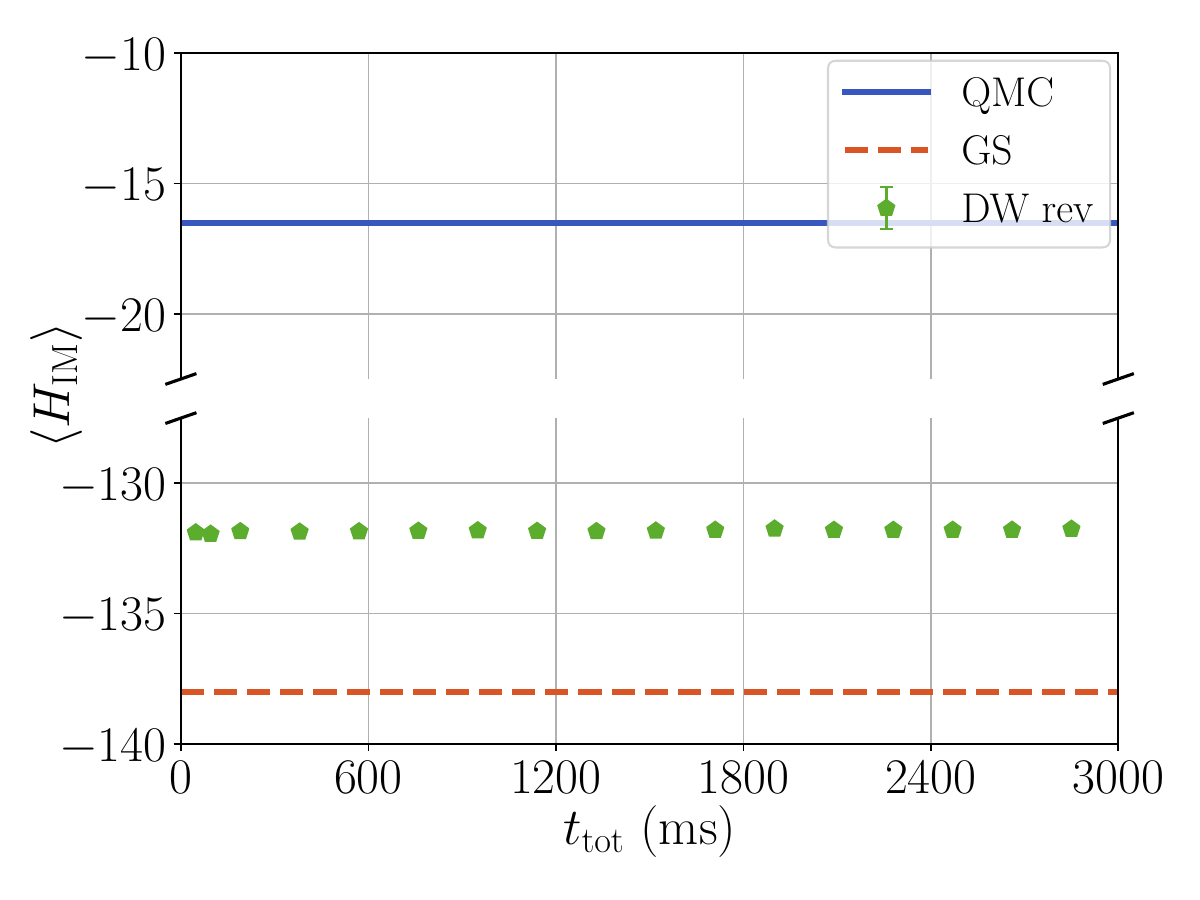}
    \par \medskip
    \caption{Expectation value for $H_{\mathrm{IM}}$ for $n=138$ and $s_p=0.2$.  Solid blue curve corresponds to QMC prediction, dashed orange curve corresponds to the Ising GS energy, and data points correspond to the experimental results using the reverse annealing protocols. Error bars correspond to the 95\% confidence interval calculated using a bootstrap over the 100 gauges.}
    \label{fig:before min gap2}
\end{figure}

\begin{figure}
  \centering
  \includegraphics[width=.95\columnwidth]{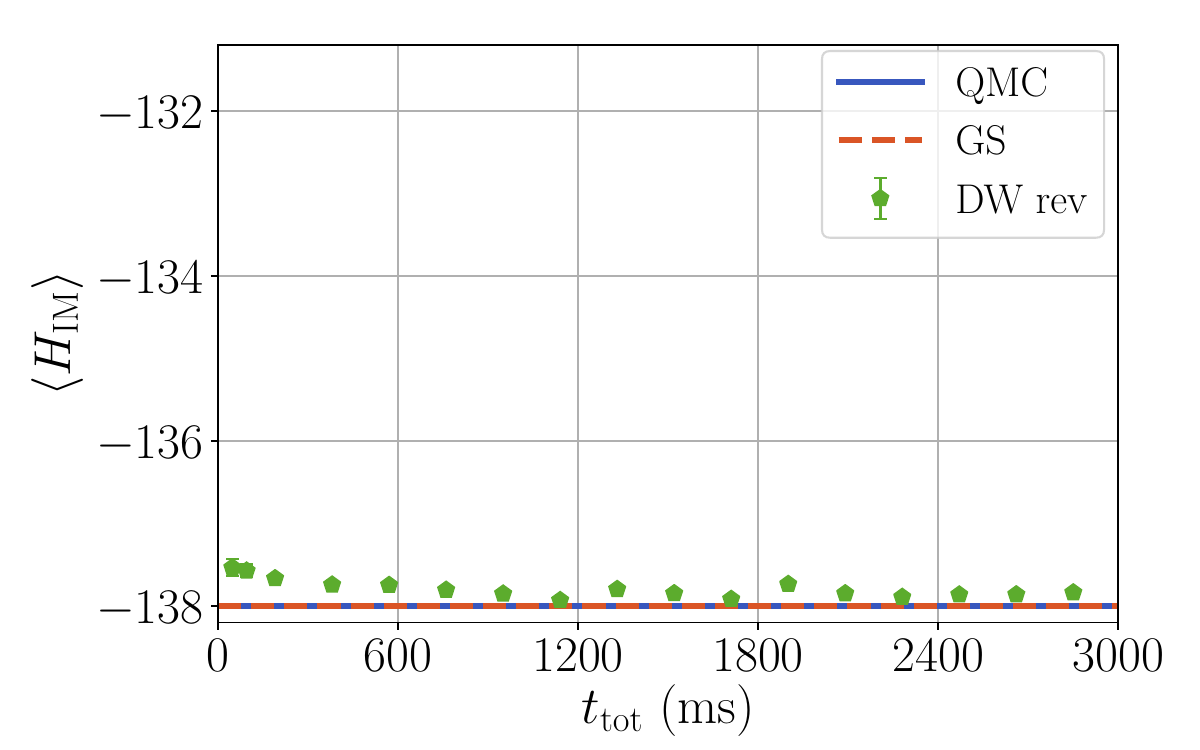}  
  \newline
  \centering
  \includegraphics[width=.95\columnwidth]{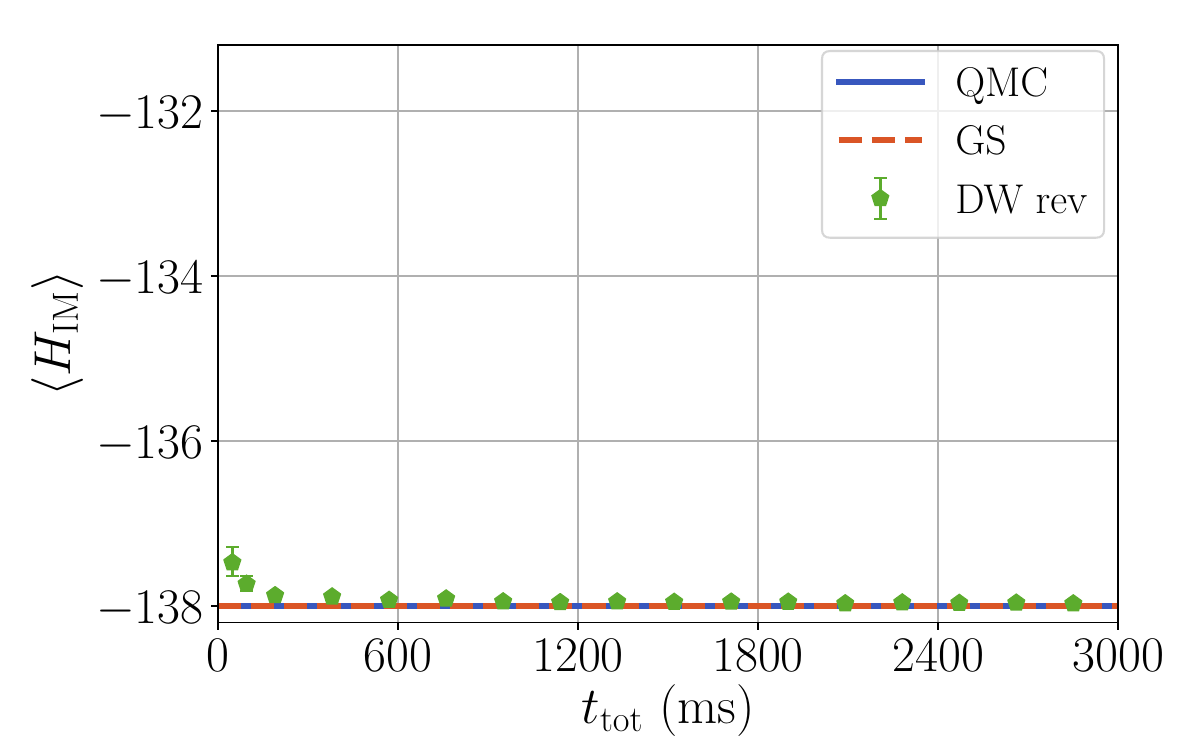} 
  \newline
  \centering
  \includegraphics[width=.95\columnwidth]{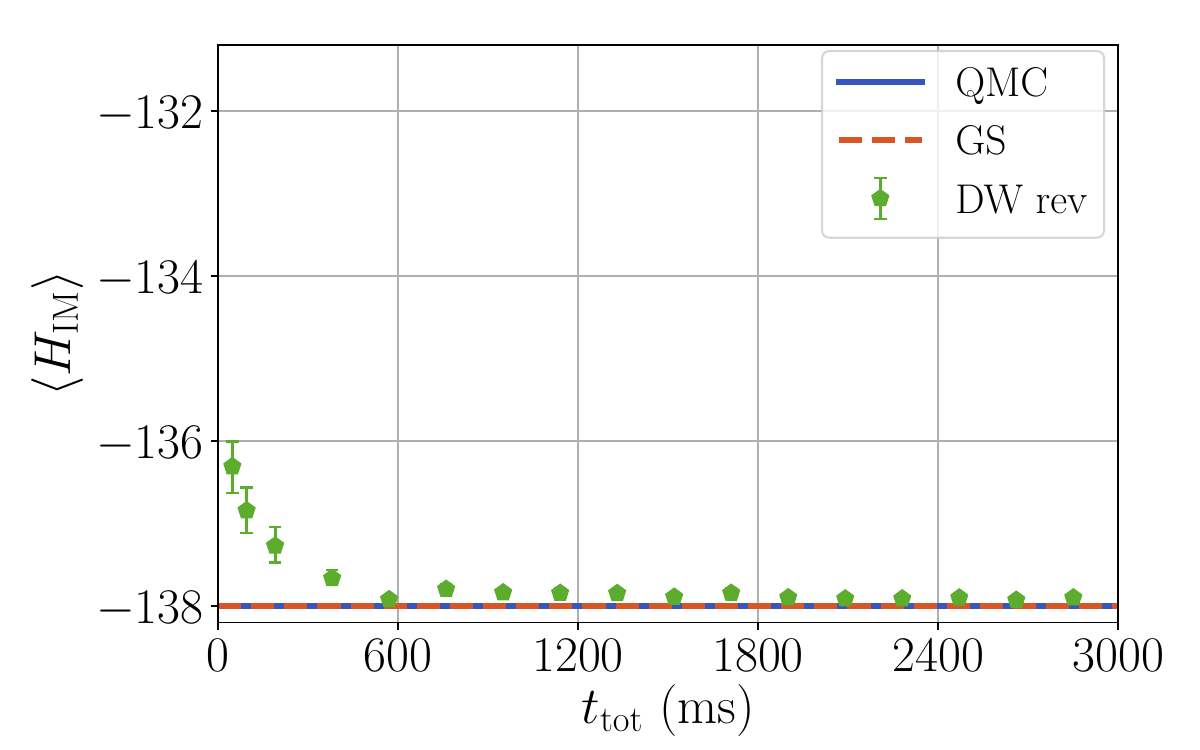}  
  \newline
  \centering
  \includegraphics[width=.95\columnwidth]{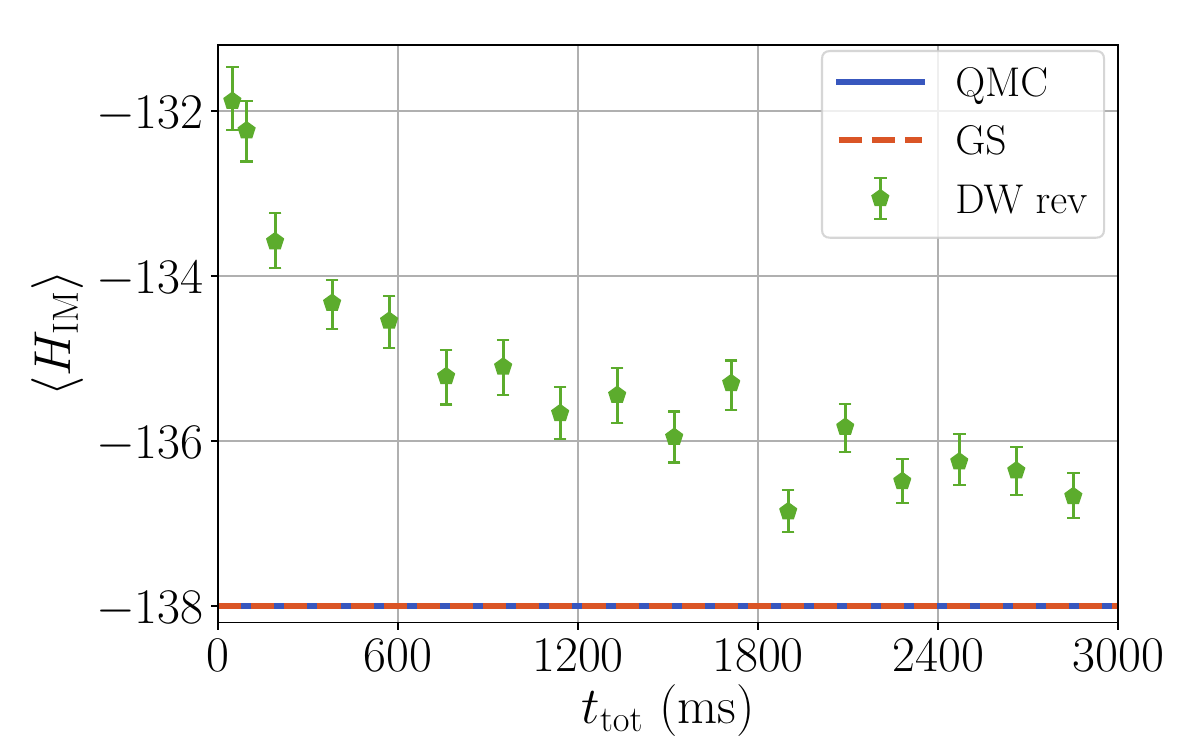}  
  \newline
\caption{From top to bottom, $\langle H_{\mathrm{IM}} \rangle$ comparison between experiment and theory for $n=138$ and $s_p=0.5, 0.6, 0.7$ and $0.8$. Solid blue curve corresponds to QMC prediction, dashed orange curve corresponds to the Ising GS energy, and data points correspond to the DW results using the  forward  annealing protocols.  Error bars correspond to the 95\% confidence interval calculated using a bootstrap over the 100 gauges.}
\label{fig:after min gap2}
\end{figure} 

For our `reverse' anneal protocol~\cite{Passarelli2020}, the system is initialized in a classical state (each qubit is either in a computational one or a computational zero state) chosen at random at $s=1$. We then anneal backwards from $s = 1$ to $s = s_p$ at a rate of $0 < \left( \frac{ds}{dt} \right)_i < 1 \mu \rm{s}^{-1}$, pause for a time $t_p$ and quench to $s=1$ at a rate of $\left( \frac{ds}{dt} \right)_f = 1 \mu \rm{s}^{-1}$. A diagram for this protocol is shown in Fig.~\ref{fig:s diagram rev}. For each subsequent anneal, we use the configuration returned from the previous anneal as our initial state. Presumably doing so instead of initializing with random state decreases the thermalization time~\cite{King2018}.  

A key difference between the forward and reverse annealing protocols is that for a sufficiently large $s_p$ the reverse annealing protocol avoids crossing the minimum gap encountered during the forward annealing protocol. This can help reduce transitions to excited states, which are more likely to happen when the spectral gap is smaller.

To allow the system to reach its steady state, we use increasing effective total time ($t_{\text{tot}}$), defined as the product of the number of anneals $n_a$ times the pause time per anneal ($t_{\text{tot}} = n_a t_p$). There are upper limits for both the time per anneal ($ \leq 2$ ms) as well as the overall total time ($ \leq 3000$ \rm{ms}). We choose a long pause time of $t_p = 1900 \mu \rm{s}$, and only vary the number of anneals $n_a$, in contrast with the forward annealing protocol where $n_a$ was fixed and a range of $t_p$ explored. Note that the maximum $t_{\text{tot}}$ is the same for both protocols.

We show the results for $s_p < s_\ast$ in Fig.~\ref{fig:before min gap2} and the results for $s_p > s_\ast$ in Fig.~\ref{fig:after min gap2}.

We see little difference between the forward and reverse anneal protocols for $s_p <s_\ast$, which is consistent with the system thermalizing rapidly in this transverse-field-dominated regime~\cite{Amin:2015qf}, but we do find some quantitative differences for $s_p > s_\ast$, which appear to become more pronounced away from $s_p =0.6$. The dependence on the annealing protocol is likely due to two different effects: i) the forward protocol must go through the minimum gap in order to reach $s= s_p$ while the reverse protocol never does, and ii) the varying initial condition of each anneal for the reverse anneal.  We see that for $s_p \leq 0.7$, the reverse annealing protocol appears to reach lower average energies than the forward annealing protocol, but for $s_p = 0.8$, the dynamics are likely too slow for it to have reached its steady state value.


\section{Frustrated Chains}
\label{App:FC}

\begin{figure}
    \centering
    \includegraphics[width=0.95\columnwidth]{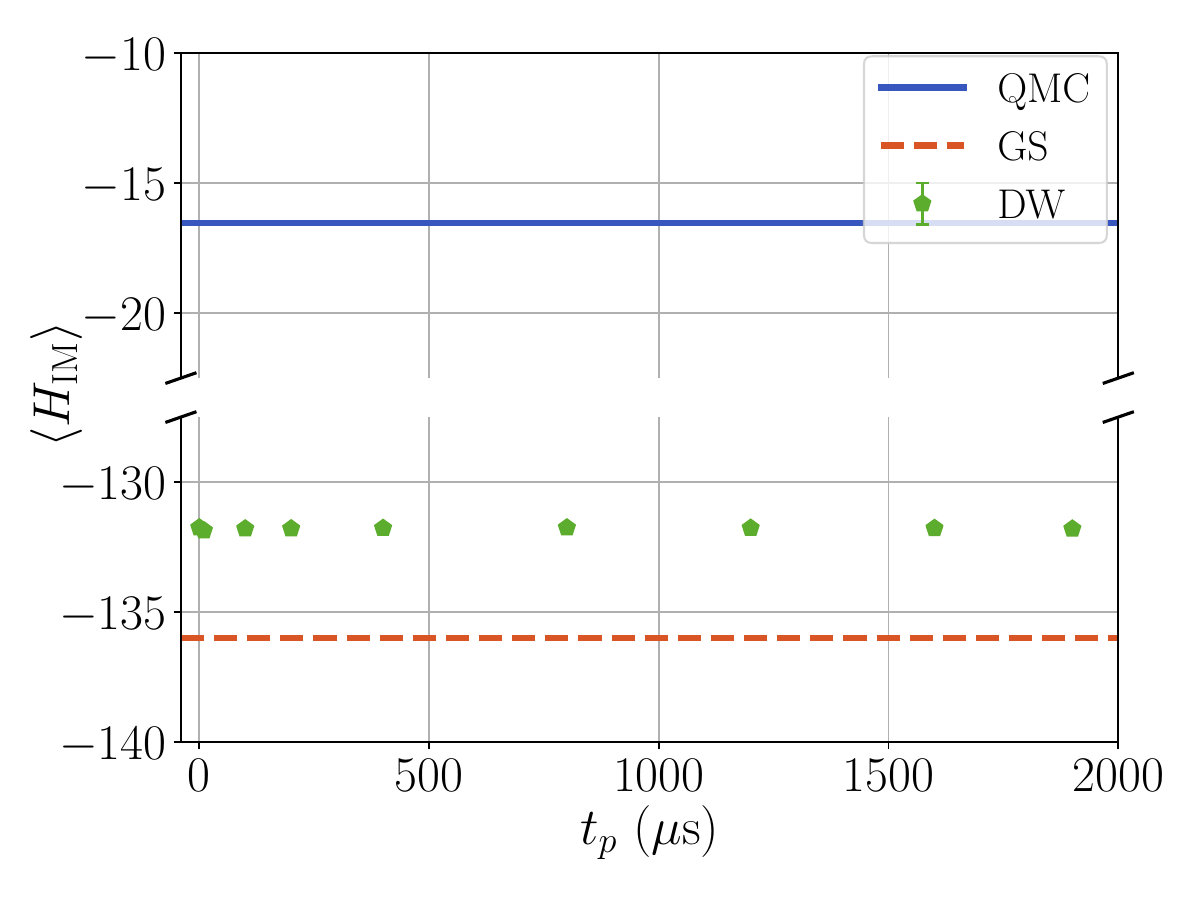}
    \caption{At $s_p=0.2$, the $n=138$ frustrated chain exhibits similar behavior to the non-frustrated case. While QMC calculations show $\langle H_{\mathrm{IM}} \rangle$ far from the GS of $H_{\mathrm{IM}}$, experimental results match what we would find at a much later point in the anneal.}
    \label{fig:frustrated}
\end{figure}

We repeated the study reported on in the main text using frustrated Ising chains. The two models are identical except that while the nonfrustrated model can be mapped to couplers having $J_{i, i+1} = -1$ $\forall$ $i$, in the frustrated case one of the couplings is antiferromagnetic and equals $+1$.
In this case, the GS configuration cannot satisfy all edges; it is $2n$-degenerate, with $E_0 = -n +2$. We find that for this model as well the results and conclusions are qualitatively very similar to the non-frustrated ones, with $\langle H_{\mathrm{IM}} \rangle$ very far from the value obtained through QMC before the minimum gap (see Fig.~\ref{fig:frustrated}).

\section{Effect of Temperature}
\label{App:ET}

\begin{figure}
    \centering
    \includegraphics[width=0.95\columnwidth]{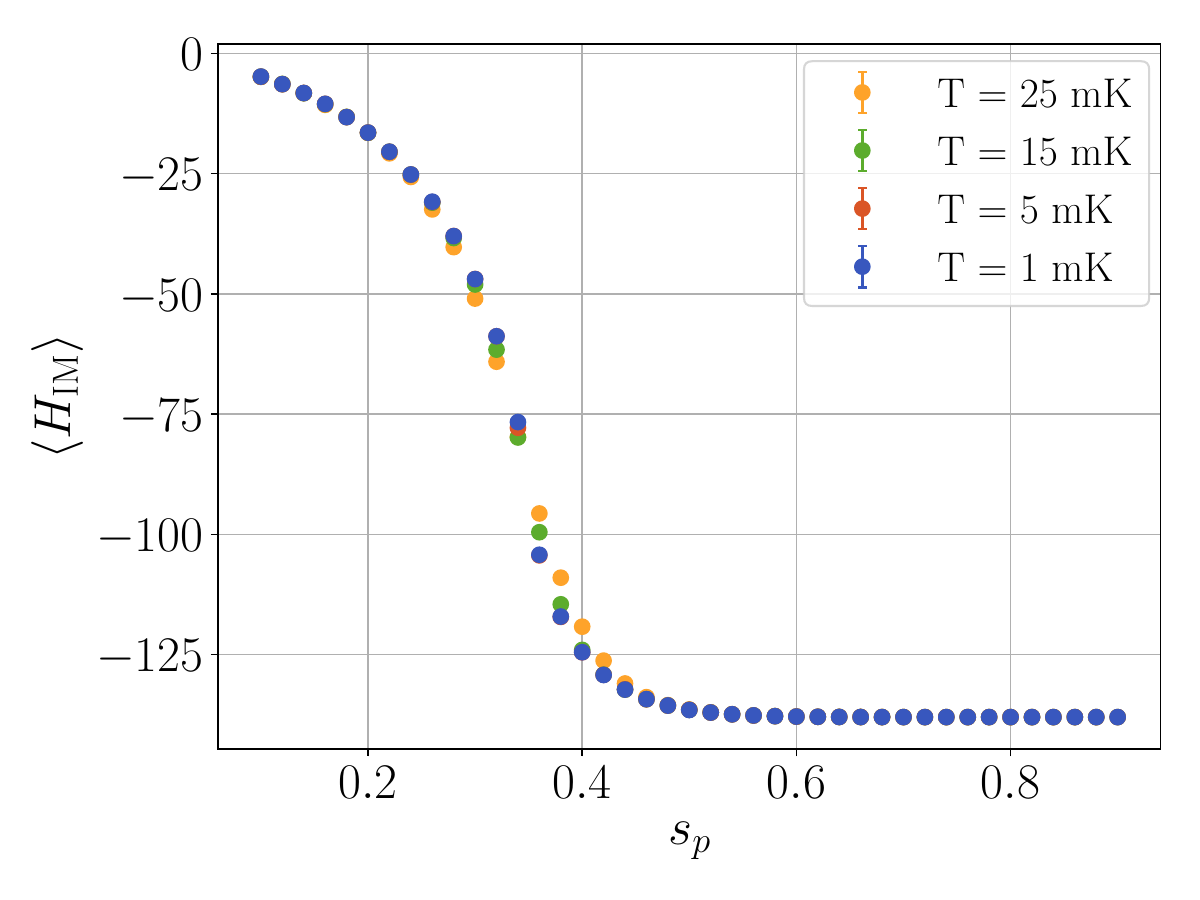}
    \caption{QMC results for the $n=138$ chain at different temperatures are shown, with very slight differences within the range of temperatures at which the annealer could potentially operate.}
    \label{fig:jw temp comp}
\end{figure}

The D-Wave 2000Q device operates at a temperature of approximately \hbox{$12$mK} which is the value we used in our QMC simulations. However, temperature is not directly measured during the experiment, and fluctuations could have an effect on our measurements~\cite{Marshall2017}. To eliminate  fluctuations in temperature as a possible source of discrepancy, we repeat the simulations for a range of temperatures, and confirm that the change in $\langle H_{\mathrm{IM}} \rangle$ is minimal and does not correspond to the behavior we observe experimentally, especially the low $\langle H_{\mathrm{IM}} \rangle$ value early in the anneal (Fig.~\ref{fig:jw temp comp}).

\section{Master equation parameters} \label{App:ME}
The master equation simulations discussed in the main text use a time-dependent Davies master equation \cite{Davies:76} that is in Lindblad form \cite{Lindblad:76}.  We have assumed identical and independent baths on each qubit, such that the master equation takes the form (with $\hbar = 1$):
\begin{eqnarray}
\frac{d}{dt} \rho(t) &=& -i \left[ {H(t)}, \rho(t) \right] + \sum_{i=1}^n \sum_{\omega} \gamma(\omega) \times \nonumber \\ 
&& \hspace{-1.5cm}  \left[ L_{\omega,i}(t) \rho(t) L_{\omega,i}(t)^\dagger  - \frac{1}{2} \left\{ L_{\omega,i}^{\dagger}(t) L_{\omega,i}(t), \rho(t) \right\} \right] \ ,
\end{eqnarray}
where the index $i$ runs over the qubits and the index $\omega$ runs over all possible Bohr frequencies of the system Hamiltonian.  The Lindblad operators are given by
\beq
L_{\omega,i}(t) = \sum_{a,b} \delta_{\omega, E_b(t) - E_a(t)} \bra{E_a(t)} \sigma^z_i \ket{E_b(t)} \ket{E_a(t)}\bra{E_b} \ ,
\eeq
where we have taken a dephasing system-bath interaction on each qubit. We assume an Ohmic oscillator bath, which then gives rise to a spectral density satisfying the KMS condition of the form:
\beq
\gamma(\omega) = \frac{2 \pi \kappa^2 \omega e^{- |\omega|/\omega_c}}{1 - e^{-\beta \omega}},
\eeq
with $\beta$ denoting the inverse-temperature.  The parameter $\kappa^2$ is a dimensionless parameter that we can tune to set the system-bath interaction strength.  For our simulations, we take it to be equal to $10^{-3}$.  The parameter $\omega_c$ is a high frequency cut-off that we set to $8 \pi$ GHz.

For the simulations used to generate the results in Fig.~\ref{fig:master eq} of the main text, we initialize the density matrix to be in the Gibbs state of the system Hamiltonian at $s_p$.  Thus, if the density matrix does not change significantly as we anneal from $s = s_p$ to $s = 1$, the density matrix populations of the computational basis states should accurately reflect those in the thermal state.

\end{document}